\documentstyle[aps,prb,preprint]{revtex}
\begin{document}
\title{Ginzburg--Landau--Gor'kov Theory of Magnetic oscillations in a type-II 
2--dimensional  Superconductor}
\author{G.~M.~Bruun, V.~Nikos Nicopoulos, N.~F.~Johnson}
\address{Department of Physics,
Clarendon Laboratory,
University of Oxford,
Oxford OX1 3PU}
\date{\today}
\maketitle

\begin{abstract}

We investigate de Haas--van Alphen (dHvA) oscillations in the mixed state of a
type-II two-dimensional superconductor within a self-consistent Gor'kov 
perturbation scheme.
Assuming that the order parameter forms a vortex lattice we can calculate the
expansion coefficients exactly to any order. We have tested  the results of the
perturbation theory to fourth and eight order against  an exact numerical
solution of the corresponding Bogoliubov-de Gennes equations. The perturbation
theory is found to describe the onset of superconductivity well close to the
transition point $H_{c2}$. Contrary to earlier calculations  by other authors
we do not find
that the perturbative scheme predicts any maximum of the dHvA-oscillations
below $H_{c2}$. Instead we obtain a substantial damping of the magnetic
oscillations in the mixed state as compared to the normal state. We have examined 
the effect of an oscillatory chemical potential due to particle conservation and the 
effect of a finite Zeeman splitting. Furthermore we have investigated the
recently debated issue of  a possibility of  a sign change of the fundamental
harmonic of the magnetic oscillations. Our theory is compared with experiment and
 we have found good agreement.
\end{abstract}
\


\section{Introduction} \label{introduction}

In recent years there has been a renewed interest in the interplay between
external magnetic fields and superconductivity in type-II superconductors. It
is well known that de Haas-van Alphen  (dHvA) oscillations are a useful tool for
probing the  Fermi surface in metals in the normal state. For type-II
superconductors the magnetic field is allowed to partially penetrate the sample
in the mixed state.  One would then expect magnetic oscillations in the mixed state to give information
about the quasi-particle dispersion and the magnetic field dependence of the
correlations in the ground state.  Magnetic oscillations in the mixed state
were observed for the first time in the layered superconductor $2H\!-\!NbSe_2$
over 20 years ago.~\cite{Graebner} More recently dHvA oscillations were
observed in the organic superconductor
$\kappa\!-\!(ET)_2Cu(NCS)_2$,\cite{Caulfield} the A15 compounds
$V_3Si$~\cite{Corcoran} and $Nb_3Sn$,~\cite{Harrison} the borocarbide
superconductor $YNi_2B_2C$,~\cite{Goll} and the high temperature
superconductors $YBaCuO$~\cite{Fowler} and $BaKBiO$~\cite{Goodrich}. These
experiments have sparked a variety of theoretical investigations,  not least
in order to understand the interplay between oscillations in the
quasi--particle spectra and the ground state condensation energy. The
transition line $H_{c2}$ between the normal state and the mixed state was shown
to exhibit weak oscillations as a function of the magnetic
field.~\cite{Rajagopal,Gruenberg} For high magnetic fields, clean samples, and very low
 temperatures $H_{c2}$ has been predicted theoretically to be a strongly oscillating 
function.~\cite{Tesanovic} The mixed state is characterized by the 
interplay between Landau level quantization due to the magnetic field, and
Cooper pair formation characteristic of superconductivity. This calls for a
theory that takes both effects into account consistently.  The theory
developed by Maki~\cite{Maki} and Stephen~\cite{Stephen} gives a simple picture
of the vortex lattice acting as an extra scattering potential on
quasi--particles thereby damping the magnetic oscillations.  The theory uses
semiclassical approximations and, crucially, fails to impose the physical
condition that the vortex lattice is the self--consistent mean field of the
Cooper pairs.  The problem simplifies when the electrons are confined to form
pairs within the same Landau level (diagonal approximation) and this case has
been treated by several authors.~\cite{1LL,Ryan} Unfortunately the diagonal
approximation ignores the fact that  the typical excitation is a
superposition of an electron and a hole in different Landau levels, but with
similar energies. This effect is strongest when the chemical potential $\mu$ is
either at a Landau level $n_f=\frac{\mu}{\hbar\omega_c}-1/2=n$  ($n$ integer)
or exactly between two Landau levels  $n_f=n+1/2$. We then have exact
degeneracy between an electron state in a Landau level $n_f+m$  and a hole in
the level $n_f-m$, when $n_f=n$, and between an electron in a level $n_f+m+1/2$
and a hole in a level $n_f-m-1/2$, when $n_f=n+1/2$, respectively.  A major
effect of the self-consistent pairing field is then to mix these two degenerate
excitations strongly.  
Following the results of the diagonal approximation Dukan \textit{et
al}.~\cite{Dukan} have focused on the consequences of a gapless portion of the
quasiparticle spectrum. The calculation, which is appropriate for low lying excitations in 3 
dimensions, is not applicable for two dimensional systems where the number of gapless points
 and their dispersion law vary strongly with the magnetic field and it does not take into 
account the oscillatory behaviour of the ground state energy as a function of the magnetic 
field.  This oscillatory behaviour of the ground state energy has been considered by 
P. Miller and B. L. Gy\"{o}rffy~\cite{Miller} in the $\Delta \gg k_bT$ limit. 
Norman \textit{et al}.~\cite{Big Mac1} have studied the problem
numerically and have linked the damping of the magnetic oscillations to the
broadening of the Landau levels due to the gap.  Recently~\cite{Maniv4} there has 
been claims based partly on Gor'kov theory and partly on an assumed simplified form 
for the quasiparticle spectrum that below a certain field $H_{inv}<H_{c2}$,
 the magnetic
 oscillations should exhibit a rapid $180^{\circ}$ phase shift.

In this paper we develop a new scheme for calculating the Gor'kov expansion terms treating 
the quantum effects of the magnetic field exactly. In addition we solve numerically the 
corresponding Bogoliubov-de Gennes (BdG) equations. Using the developed formalism 
we study the magnetic oscillations in the mixed state of a type II superconductor. We are 
working in two dimensions since 
many organic metals are known to show almost perfect 2D behaviour. 
Exploiting the symmetry of the magnetic translation group of the vortex lattice
we have been able to calculate the expansion coefficients in the Gor'kov theory 
exactly to any order making no restriction on the energy of the 
center-of-mass of the Cooper pairs. Self-consistency within this approach then
 transforms to the much simpler problem of minimising a polynomial of a finite
 number of variables. This allows us to develop an analytical theory 
for the thermodynamic potential and thus for the magnetic oscillations close to 
$H_{c2}$ which contains no approximations 
apart from the assumption of a small order parameter. This establishes a rigorous basis for 
our theory, compared with earlier attempts. It turns out to be crucial to 
determine the order parameter self-consistently since its oscillatory behaviour  when
 the magnetic field varies is the cause of the damping of the dHvA oscillations. 
We find that the dHvA oscillations are damped in the mixed
state as compared to the normal state, in agreement with what is observed
experimentally. This is due to the fact that the contribution from the
superconducting order parameter to the magnetic oscillations partly cancels the
contribution from the normal grand potential.  The superconducting order
parameter itself is an oscillating function of the magnetic field, with local
maxima occurring whenever we have a Landau level at the chemical potential since
electrons can then form Cooper pairs without any cost in kinetic energy.  This
is the simple physical picture of  the damping emerging from our formalism and it 
complements the interpretation given by P. Miller and B. L. Gy\"{o}rffy~\cite{Miller} and
 Norman \textit{et al}.~\cite{Big Mac1} When many Landau levels participate in the pairing we have simplified the expressions 
for the expansion parameters. This makes it possible to give fairly simple analytical 
expressions for the rate of damping af the dHvA oscillations close to the transition line that 
may prove useful when fitting experimental data.

 A similar approach has been taken by Maniv \textit{et al}.~\cite{Maniv1} Using the
semiclassical and various other approximations, they calculate the Gor'kov expansion 
coefficients for a 2D metal to fourth order in $\Delta(\mathbf{r})$  when the motion of
 the centers of mass of the Cooper pairs is restricted to the lowest Landau level. However, they 
obtain~\cite{Maniv2} that the magnitude of the magnetic oscillations exhibit a maximum below
 $H_{c2}$. This is contradicted by our exact calculation of the expansion coefficients and 
also by our numerical solution to the BdG-equations.

Recently it has been suggested that the degeneracy of the Landau levels should give rise to 
non-perturbative terms in the expansion of the grand potential thereby making the traditional 
Gor'kov theory invalid.~\cite{Bahcall}  We have tested our perturbative theory carefully against an exact
numerical  solution of the BdG-equations and we do not find any of the predicted non-perturbative
 effects. The theory based on the Gor'kov expansion agrees very well 
with the exact solution if we are not too far below $H_{c2}$. It is essentially a 
high temperature expansion in the sense that is an asymptotic series as long as 
the change in the quasiparticle levels as compared to the normal state is not larger than 
 $\sim O(k_bT)$.~\cite{Bruun}

In a two-dimensional metal the chemical potential is an oscillatory function of the magnetic
 field when the number of particles $N$ is fixed. When higher harmonics are important (i.e. low 
temperatures and clean samples) the dHvA 
signal in the normal state for fixed $N$ look qualitatively different from the case when the chemical 
potential is fixed. It is of interest to see what consequences 
this difference has for the magnetic oscillations in the mixed state. 
Examination of the dHvA 
oscillations in the mixed state in the two cases yields that the
 superconducting order for fixed number of particles reduces the oscillations in the chemical potential 
and that the dHvA oscillations are essentially the same in the two cases apart from a narrow region 
close to $H_{c2}$. Specifically, the rate of damping of the magnetic oscillations is the same when 
the number of particles is constant and when the chemical potential is constant.

Since the contribution to the magnetic oscillations
from the  condensation energy is in antiphase with the contribution from
the normal grand potential, it has been suggested~\cite{Maniv4} that this will 
result in  a sign change of the fundamental harmonic of the dHvA oscillations for 
 $H\leq H_{inv}<H_{c2}$. This would happen if the
superconducting contribution were to overwhelm the contribution from the normal
grand potential deep enough into the superconducting state. Based on an 
approximate evaluation of the Gor'kov expansion parameters one can calculate  
an expression for $H_{inv}$.~\cite{Maniv4}  Using
our  expressions for the damping, we are able to predict that within
the region of validity of the perturbative scheme this effect will not occur. 
Hence there is no theoretical reason, within perturbation theory, to expect
inversion of the magnetic oscillations.  This result agrees with the lack of
experimental observation of such an effect. It also agrees with our exact numerical 
solutions to the BdG equations which show a complete suppression of the magnetic oscillations deep enough into the mixed state.~\cite{Big Mac3}

Although there are currently experimental uncertainties about the value of
$H_{c2}$ in the organic superconductors, a comparison with experimental results
 for the quasi 2D superconductor $\kappa\!-\!(ET)_2Cu(NCS)_2$ yields good agreement
 between theory and experiment.

The outline of our paper is as follows. Sec.\ \ref{eiv} sets up the formalism
for describing the vortex state using both the Bogoliubov-de Gennes equations
and perturbation theory.  In Sec.\ \ref{numerical} we compare the results of
the perturbation theory with the  exact numerical solution.  The damping of the
magnetic oscillations is discussed in Sec.\ \ref{damping}. We give a physical
interpretation of the damping. The effect of a finite Zeeman splitting term is 
discussed and the case of a conserved number of particles as opposed to a conserved 
chemical potential is considered. Using approximate expressions for the damping 
parameters we are able to give a simple analytical expression for the rate of damping of the dHvA 
oscillations close to $H_{c2}$. The spin dependence and the temperature dependence of the
oscillations can then be extracted. We then examine the validity of the arguments leading to a sign
 change of the first harmonic of the dHvA oscillations. In Sec.\ \ref{comparison} we compare 
our analytical theory with experimental results. Finally we summarize our results
in Sec.\ \ref{conclusion}.

\section{Electrons in the Vortex State}\label{eiv}

\subsection{General representation and BdG-equations}

 We consider a pure 2D electron gas in the $x-y$ plane with a perpendicular
magnetic field H along the $z$-axis. In the Landau gauge, ${\bf A}=(0,Hx,0) $,
the single--particle  eigenstates can be chosen to be
\begin{equation}\phi_{N,k}({\bf r})=\frac{1}{\sqrt{L_y}}e^{-iky} 
\phi_N(\frac{x-kl^2}{l})\end{equation} where $\phi_N(x)=(2^N
N!\sqrt{\pi}l)^{-1/2} H_N(x)e^{-\frac{1}{2}x^2} $ with $H_N$ being a Hermite
polynomial of order N, and $l^2=\hbar c/eH$ is the magnetic length. The size of
the system is $L_x\times L_y$. Band structure effects are assumed to be
adequately described employing an electron effective mass $m^*$. The B-field is
taken to be uniform within the sample thereby ignoring the partial screening by
the supercurrents. This approximation holds for strong type-II superconductors
($\kappa \gg 1$) such as the organics, where the penetration depth is much
larger than the coherence length.

In the mixed state of a conventional type II superconductor the order parameter forms a 
vortex lattice. It is therefore advantageous to use a basis set which incorporates this 
symmetry. 
We have chosen to use the following set of functions introduced 
by Norman \textit{et al}:~\cite{Big Mac1}
 \begin{equation} \phi_{N {\bf k}}({\bf r})=
\sqrt{\frac{a_x}{L_x}}\sum_t e^{ik_x a_xt} e^{i\pi t^2/4}\phi _{N,-k_y
+ta_x/l^2}( {\bf r})  \end{equation}
 where $k_x \in [0, \frac{2 \pi}{a_x}[$
with $\Delta k_x= \frac{2\pi}{L_x}$ and $k_y \in [0,a_x/l^2[$ with $\Delta
k_y=\frac{2 \pi}{L_y}$ define the magnetic Brillouin zone (MBZ). The symmetry 
of the order parameter restricts the pairing to be between electrons with quantum numbers 
\textbf{k} and -\textbf{k}.~\cite{1LL} By adjusting $a_x$ we can obtain both a triangular
(\mbox{$a_x=l(\sqrt{3} \pi /2)^{1/2}$} ) and a square vortex lattice
(\mbox{$a_x=l(\pi /2)^{1/2}$} ). Throughout this paper we choose to work with
the triangular lattice since we expect the free energy to be minimized by this
symmetry (except possibly in the re-entrant regime).~\cite{Big Mac1}
We are using  mean field BCS-theory with a smooth cutoff in the interaction around the 
Fermi surface, applicable for weak-coupling superconductors. The mean-field Hamiltonian is:
\begin{eqnarray} \label{hamilton}\hat{H}&=&\hat{H_0}+\hat{H_1}
\nonumber \\ \hat{H_0}&=&\int d{\bf r}\, \psi_{\sigma}^{\dagger}({\bf r})
\left( \frac{({\bf p}-\frac{e}{c}{\bf A})^2}{2m} -\mu\right)\psi _{\sigma}({\bf
r}) \nonumber \\ \hat{H_1}&=&\sum _{\stackrel{N M}{{\bf k}}}\int d{\bf
r}\,[\Delta({\bf r})w(N)w(M)\phi_{M{\bf k}}^ {\ast}({\bf r}) \phi_{N-{\bf
k}}^{\ast}({\bf r})\hat{a}_{M {\bf k}\uparrow}^{\dagger} \hat{a}_{N -{\bf k}
\downarrow}^{\dagger}+c.c.] \end{eqnarray}
where the order parameter is defined as:
\begin{equation} \Delta( {\mathbf{r}} )\equiv
V\sum_{\stackrel{NM}{\mathbf{k}}} w(N)w(M)\phi_{N\mathbf{k}} ({\mathbf{r}})
\phi_{M-\mathbf{ k}}({\mathbf{r}})<\!
a_{N{\mathbf{k}}\uparrow}a_{M-\mathbf{k}{\downarrow}}\!> \end{equation}
This differs from the conventional BCS-hamiltonian since we have 
introduced the weight function $w(n)$. It is necessary to have a smooth cutoff 
in the pairing interaction since we otherwise would get non-physical effects 
arising from Landau levels abruptly entering or leaving the pairing region. 
 The weight function $w(n)$ is chosen to be Gaussian i.e.  
$w(N) \propto e^{-(\xi_N /0.5\hbar\omega_D)^2}$
where $ \omega_D$ is the pairing width and $\xi_{N} =(N+1/2 )\hbar \omega_c-\mu$. 
This approach was introduced by Norman 
\textit{et al}~\cite{Big Mac1} although they used a different weighting function.
It should be noted that the above slightly unconventional definition of the order 
parameter is necessary. Otherwise the self-consistency condition is not
equivalent to minimising the grand potential $\Omega$ with respect to $\Delta
(\mathbf{r})$ (i.e. $\frac{\delta \Omega} {\delta \Delta(\mathbf{r})}=0$).
 In the vortex lattice case the order parameter can be characterized by a 
finite number of parameters $\Delta _j$~\cite{Big Mac1} 
\begin{equation}
\Delta({\bf r})=\frac{Va_x}{\sqrt{lL_y}L_x} \sum _j \Delta_j \sum _g e^{i\pi
g^2/2}\phi_{j,\sqrt{2} ga_x}(\sqrt{2} {\bf r}) \end{equation}
The $\Delta_j$'s are determined selfconsistently as explained in reference 17. 
 Assuming not only translational but also six fold rotational
symmetry of  $|\Delta({\bf r})|$ gives the restriction~\cite{Ryan}
$j=0,6,12,\ldots$ where $j\leq2N_{\max}$. $N_{\max}$ is the highest Landau
level participating in the pairing.   Using
the above transformation  the corresponding BdG-equations split into a set of equations for 
each ${\mathbf{k}}$ and they can be solved numerically.  
Norman \textit{et al}.~\cite{Big Mac1} have carried out  an
extensive numerical investigation of the quasiparticle spectrum and the
magnetic oscillations in the superconducting state. We have developed a
similar numerical scheme to solve the BdG-equations. In this way we are able to
check our analytical results against an exact numerical solution.

\subsection{Perturbative expansion of the grand potential}
Since we are interested in the region near $H_{c2}$ where the order parameter is small, it is natural to 
consider the Gor'kov expansion of the grand potential.  This can be done either through the equation of motion approach originally used or by using the grand partition function for  the symmetry-broken self-consistent 
Hamiltonian:
\begin{eqnarray} {\cal Z} =\int {\cal D} (\psi_{\sigma}^{\ast }({\bf r},\tau ) \psi_{\sigma}
({\bf r},\tau )) e^{-\int _0 ^{\beta} d\tau \int d{\bf r} {\cal L }({\bf r},\tau)} \nonumber \\
{\cal L}({\bf r},\tau)= \psi_{\sigma}^{\ast}({\bf r},\tau ) \left(  \partial _{\tau} +
\frac{({\bf p}-\frac{e}{c}{\bf A})^2}{2m}    -\mu\right)\psi_{\sigma}({\bf r},\tau) 
\nonumber \\ - \left[ \Delta({\bf r}) \tilde{\psi}_{\downarrow} ^{\ast}({\bf r}, \tau) 
\tilde{\psi}_{\uparrow}^{\ast }({\bf r}, \tau)+c.c- \frac{1}{V} |\Delta ({\bf r})|^2 \right]
\end{eqnarray}
where ${\cal D } (\psi_{\sigma}^{\ast }({\bf r},\tau ) \psi_{\sigma}({\bf r},\tau ))$ denotes functional 
integration over Grassman variables.
We have defined $\tilde{\psi}_{\sigma}({\bf r})=\sum_{n,{\mathbf{k}}}w(n)\phi_{n,{\mathbf{k}}}
({\mathbf{r}})a_{n,{\mathbf{k}}\sigma}$.
The $\frac{1}{V}|\Delta({\mathbf{r}})|^2$ term  corrects for the double counting of the interaction energy in the Hartree--Fock approximation.
Expanding the grand potential $\Omega=-\frac{1}{\beta} \ln {\cal Z}$ in powers of $\Delta ({\bf r}) $ we
 obtain to eighth order
 \begin{equation}\Omega _S -\Omega_N =\Omega _2+\Omega_4 +\Omega_6 +\Omega_8\end{equation} 
where
\begin{eqnarray} \Omega _2= \frac{1}{V} \int d{\bf r} | \Delta ({\bf r}) |^2 -\frac{1}{\beta} \int d{\bf r}_1 
d{\bf r}_2 \Delta({\bf r}_1) \Delta ^{\ast} ({\bf r}_2) K_2 ({\bf r}_1 ,{\bf r}_2) \nonumber \\ \Omega_4=
\frac{1}{2\beta} \int d{\bf r}_1 \ldots d{\bf r}_4 K_4 ({\bf r}_1,{\bf r}_2,{\bf r}_3,{\bf r}_4) \Delta({\bf r}_1) 
\Delta({\bf r}_2) \Delta^{\ast}({\bf r}_3) \Delta^{\ast} ({\bf r}_4)\nonumber \end{eqnarray}
\begin{eqnarray}\Omega_6=-\frac{1}{3\beta} \int d{\bf r}_1 \ldots d{\bf r}_6 K_6({\bf r}_1,\ldots,
{\bf r}_6)\Delta({\bf r}_1)\Delta({\bf r}_2) \Delta({\bf r}_3) \Delta^{\ast}({\bf r}_4) \Delta^{\ast}
({\bf r}_5) \Delta ^{\ast}({\bf r}_6) \nonumber \\ \Omega_8=\frac{1}{4\beta} \int d{\bf r}_1 \ldots d{\bf r}
_8 K_8({\bf r}_1,\ldots,{\bf r}_8) \Delta({\bf r}_1)\Delta({\bf r}_2) \Delta({\bf r}_3) \Delta({\bf r}_4)
 \Delta^{\ast}({\bf r}_5) \Delta^{\ast}({\bf r}_6) \Delta ^{\ast}({\bf r}_7) \Delta ^{\ast}({\bf r}_8)
\end{eqnarray}
 The kernels are given by
\begin{eqnarray} K_2({\bf r}_1,{\bf r}_2)=\frac{1}{\hbar ^2} \sum _{\nu} \tilde{G}^{\circ}_{\uparrow}
({\bf r}_2,{\bf r}_1,-\omega_{\nu}) \tilde{G}^{\circ}_{\downarrow} ({\bf r}_2,{\bf r}_1,\omega_{\nu}) 
\nonumber \\ K_4 ({\bf r}_1,{\bf r}_2,{\bf r}_3,{\bf r}_4)=\frac{1}{\hbar ^4}\sum _{\nu} \tilde{G}^{\circ}
_{\downarrow}({\bf r}_4,{\bf r}_1,\omega_{\nu}) \tilde{G}^{\circ}_{\uparrow} ({\bf r}_3,{\bf r}_1,-
\omega_{\nu}) \tilde{G}^{\circ}_{\downarrow}({\bf r}_3,{\bf r}_2,\omega_{\nu}) \tilde{G}^{\circ}_
{\uparrow} ({\bf r}_4,{\bf r}_2,-\omega_{\nu}) \nonumber 
\\ K_6({\bf r}_1,\ldots,{\bf r}_6)=\frac{1}{\hbar ^6}\sum_{\nu}\tilde{G}^{\circ}_{\downarrow}({\bf r}_6,
{\bf r}_1,\omega_{\nu}) \tilde{G}^{\circ}_{\uparrow} ({\bf r}_5,{\bf r}_1,-\omega_{\nu}) \tilde{G}^{\circ}_
{\downarrow}({\bf r}_5,{\bf r}_2,\omega_{\nu}) \nonumber \\ \times \tilde{G}^{\circ}_{\uparrow} 
({\bf r}_4,{\bf r}_2,-\omega_{\nu}) \tilde{G}^{\circ}_{\downarrow} ({\bf r}_4,{\bf r}_3,\omega_{\nu}) 
\tilde{G}^{\circ}_{\uparrow} ({\bf r}_6,{\bf r}_3,-\omega_{\nu}) \nonumber \\
 K_8({\bf r}_1,\ldots,{\bf r}_8)=\frac{1}{\hbar ^8}\sum_{\nu}\tilde{G}^{\circ}_{\downarrow}({\bf r}_6,
{\bf r}_1,\omega_{\nu}) \tilde{G}^{\circ}_{\uparrow} ({\bf r}_7,{\bf r}_1,-\omega_{\nu}) \tilde{G}^{\circ}
_{\downarrow}({\bf r}_7,{\bf r}_2,\omega_{\nu}) \tilde{G}^{\circ}_{\uparrow} ({\bf r}_8,{\bf r}_2,-
\omega_{\nu}) \nonumber 
\\ \times \tilde{G}^{\circ}_{\downarrow} ({\bf r}_5,{\bf r}_3,\omega_{\nu})\tilde{G}^{\circ}_{\uparrow} 
({\bf r}_6,{\bf r}_3,-\omega_{\nu}) \tilde{G}^{\circ}_{\downarrow} ({\bf r}_8,{\bf r}_4,\omega_{\nu})
 \tilde{G}^{\circ}_{\uparrow} ({\bf r}_5,{\bf r}_4,-\omega_{\nu})  \end{eqnarray}
 and $\omega_{\nu}=(2\nu+1)\pi k_b T/\hbar$ are the Matsubara frequencies. Maniv 
\textit{et al}.~\cite{Maniv1} have calculated 
the expansion up to fourth order in $\Delta( {\textbf r}) $ using essentially  semiclassical  approximations. 
They used a  variational form of the order parameter which has no symmetry built in initially 
but restricts the electrons to condense in the lowest center-of-mass Landau level
 ($\Delta_{j\neq 0}=0$). As will be shown below, this restricion introduces no serious error within in the 
region of interest in the phase diagram.
 Since it is known~\cite{Big Mac1} that the triangular lattice is the minimal 
energy configuration (except for the re--entrant regime) we have exploited this symmetry  to calculate these expansion terms  exactly.
 Because we are using a smooth pairing cutoff in our  Hamiltonian we have, instead of the Green's function for the 
normal state $G^{\circ}_{\sigma}({\bf r}_2,{\bf r}_1,\omega_{\nu})$, the following function in our kernels:
\begin{equation} \tilde{G}^{\circ}_{\sigma}({\bf r}_2,{\bf r}_1,\omega_{\nu})=\sum _{n{\bf k}}
\frac{\phi_{n{\bf k}}({\bf r}_2) \phi_{n{\bf k}}^{\ast}({\bf r}_1)} {i\omega_{\nu}-\xi_{n\sigma}/\hbar}
w^2(n)\end{equation} 
where $\xi_{n\sigma} =\xi_n+gm^*\sigma /2m_0 \hbar \omega_c$. The only difference from 
the Green's 
function for the normal state is that we have included the weight functions $w(n)$ in the sum. 
Using the symmetry of the vortex lattice the integrals can be solved. 
We have to fourth order
 \begin{eqnarray} \label {alpha}
 \Omega_2 &=& \frac{Va_x}{\sqrt{2}lL_xL_y} \sum _j \left[ 1-\frac{V}{4\pi l^2} \sum _{n_1,n_2}
 {B_j^{n_1\,n_2}}^2w^2(n_1)w^2(n_2) \frac{\tanh (\beta \xi_{n1 \downarrow}/2)+
 \tanh (\beta \xi_{n2 \uparrow}/2)}{2(\xi_{n_1 \downarrow}+\xi_{n_2 \uparrow}) }
\right]\Delta_j ^2 \nonumber \\ &=& \sum_j \alpha_j \Delta_j^2\end{eqnarray}
and
\begin{eqnarray}  \label{gamma} \Omega_4 &=& \frac{V^4 a_x^4}{8L_x^4L_y^4l^4} 
\sum_{n_1 \ldots n_4}w^2(n_1)w^2(n_2)w^2(n_3)w^2(n_4) f(n_1,n_2,n_3,n_4) \nonumber \\ 
&\times& \sum_{j_1 \ldots j_4}B_{j_1}^{n_1n_4}B_{j_2}^{n_3 n_2}  B_{j_3}^{n_1n_2}B_{j_4}^{n_3 n_4}
   \Xi^{n_1+n_4-j_1,n_2+n_3-j_2}_{n_1+n_2-j_3,n_3+n_4-j_4} \Delta_{j_1} \Delta_{j_2} 
\Delta_{j_3} \Delta_{j_4} \nonumber \\ &=& \sum_{j_1 \ldots j_4} \gamma_{j_1 \ldots j_4} 
\Delta_{j1} \Delta_{j_2} \Delta_{j_3} \Delta_{j_4}\end{eqnarray}
 where 
\begin{eqnarray} \label{fsum1} f(n_1,n_2,n_3,n_4)=\frac{1}{\beta} \sum_{\nu}\left[ (-i\hbar \omega_
{\nu}-\xi_{n_1 \downarrow}) (i\hbar \omega_{\nu}-\xi_{n_2 \uparrow})(-i\hbar \omega_{\nu}-
\xi_{n_3 \downarrow}) (i\hbar \omega_{\nu}-\xi_{n_4 \uparrow})\right]^{-1}  \end{eqnarray} 
and 
\begin{eqnarray} \label{xi} \Xi_{j_3,j_4}^{j_1,j_2}=\frac{L_xL_y}{4\pi a_x} 
\sum_j B_j^{j_1 j_2}B_j^{j_3 j_4} \nonumber \sum_{h_1 h_2}e^{-i\pi (h_1^2-h_2^2)}\left[\phi_{j_1+j_2-j}
(2h_1a_x)\phi_{j_3+j_4-j}(2h_2a_x)+\right. \nonumber \\ \left. e^{-i\pi (h_1-h_2)} \phi_{j_1+j_2-j}
(2h_1a_x+a_x)\phi_{j_3+j_4-j} (2h_2a_x+a_x)\right]  \end{eqnarray}
The coeffiecient $B_j^{N M}$ is defined as
\begin{equation} B_j ^{N,M}\equiv \left( \frac {j!(N+M-j)!N!M!} {2^{N+M}} \right )^{1/2} 
\sum_{m=\max(0,j-N)}^{\min(j,M)} \frac{(-1)^{M-m}} {(j-m)!(N+m-j)!(M-m)!m!} \end{equation} 

The sums over states above are restricted to Landau levels lying within the pairing width around the 
chemical potential. Using the standard method of evaluating Matsubara sums by contour integration we obtain
\begin{eqnarray} f(n_1,n_2,n_3,n_4)= \left[(e^{-\beta \xi_{n_1 \downarrow}}+1) (\xi_{n_1 \downarrow}
+\xi_{n_2 \uparrow})(-\xi_{n_1 \downarrow}+\xi_{n_3 \downarrow})(\xi_{n_1 \downarrow}+
\xi_{n_4 \uparrow})\right]^{-1} +\nonumber \\  \left[(e^{\beta \xi_{n_2 \uparrow}}+1) (\xi_{n_2 
\uparrow}+\xi_{n_1 \downarrow}) (\xi_{n_2 \uparrow}+\xi_{n_3 \downarrow})(\xi_{n_2 \uparrow}-
\xi_{n_4 \uparrow}) \right]^{-1}+ \nonumber \\  \left[(e^{-\beta \xi_{n_3 \downarrow}}+1) (-\xi_{n_3 
\downarrow}+\xi_{n_1 \downarrow})(\xi_{n_3 \downarrow}+\xi_{n_2 \uparrow})(\xi_{n_3 
\downarrow}+\xi_{n_4 \uparrow})\right]^{-1} +\nonumber \\ \left[(e^{\beta \xi_{n_4 \uparrow}}+1) 
(\xi_{n_4 \uparrow}+\xi_{n_1 \downarrow})(\xi_{n_4 \uparrow}-\xi_{n_2 \uparrow})(\xi_{n_4 
\uparrow}+\xi_{n_3 \uparrow}) \right]^{-1} \label{matsusum}  \end{eqnarray}

  The second order term
 $\Omega_2$ which determines the $H_{c2}$ line agrees, apart from the inclusion of the 
 weight 
function, with the result of MacDonald \textit{et al}.~\cite{Big Mac2} and Rajagopal and Ryan.~\cite{Rajo1} 
 The six and eighth order terms $\Omega_6$ and $\Omega_8$ can also be calculated and they are given in 
appendix \ref{68terms}. 
We get the form:
\begin{eqnarray}\ \Omega_S-\Omega_N=\sum_j \alpha_j(T,H) \Delta_j^2+ \sum_{j_1 \ldots j_4}
 \gamma_{j_1 \ldots j_4}(T,H) \Delta_{j_1} \cdots \Delta_{j_4}+ \nonumber \\ \sum_{j_1 \ldots j_6} 
\kappa_{j_1 \ldots j_6}(T,H) \Delta_{j_1} \cdots \Delta{j_6}+\sum_{j_1 \ldots j_8} \eta_{j_1 \ldots j_8}
(T,H) \Delta_{j_1} \cdots \Delta_{j_8} \label{expan}\end{eqnarray}

Thus we have derived the exact quantum mechanical expressions for the
expansion coefficients for $\Omega_S-\Omega_N$ up to eighth order  assuming a
vortex lattice. We have not yet restricted the
electrons to form pairs with the lowest possible center-of-mass energy ($j=0$). 
The result is a multidimensional polynomial in $\Delta_j$. Going to eighth order permits 
us to check the convergence properties of the series. We could in
principle calculate the expansion coefficients to any order but, as usual, the
algebra gets more  cumbersome with increasing order, and the minimization
condition cannot be solved analytically for such high orders.  

\subsection{Self-consistency and minimization of $\Omega_S$} \label{parameters}
The self-consistent determination of  \mbox{$\Delta ({\mathbf{ r}}) \equiv V<\!\psi_{\uparrow}({\bf r}) 
\psi_{\downarrow} ({\bf r})\!>$} is equivalent to minimising the grand potential with respect to 
$\Delta({\bf r})$.~\cite{Eilenberger} In the above  formulation, which takes into 
account the spatial symmetry of the order parameter, this reduces to minimising our multi-dimensional
 polynomial with respect to $\Delta_j$. Although this is a standard numerical problem it is necessary to make 
further aproximations in order to obtain simple analytical results. The instability towards 
superconductivity is determined by the sign 
of the expansion coefficients $\alpha_j$. Above $H_{c2}$ we have $\alpha_{j}>0$ for all $j$. The transition 
to the mixed vortex state occurs when one of the $\alpha_j$'s becomes negative. The system can then lower
 its energy by making the corresponding $\Delta_j$ nonzero. It has been shown that the instability occurs first in the
 j=0 channel.~\cite{Big Mac2} So we have $\alpha_0 <0$ and $\alpha_{j\neq 0}>0$ for $H\,
\raisebox{-0.4ex}{$\stackrel {<}{\sim}$}\,H_{c2}$ and therefore \mbox{$\Delta_0 \gg \Delta_{j\neq 0}$}. 
We can then make 
the approximation $\Delta_{j\neq 0}=0$, i.e only consider condensation into
pairs with lowest Landau level center--of--mass motion. We have checked this approximation 
by solving the BdG equation numerically when $\Delta_{j\neq0}=0$ and when all the 
$\Delta_j$'s are non-zero. In the region of interest there is essentially no difference between 
the two solutions thus justifying our approximation.

The  grand potential now has the Landau form 
\begin{equation} \label{Landau form}
\alpha \Delta_0^2+\gamma \Delta_0^4 +\kappa 
\Delta_0^6 +\eta \Delta_0^8, 
\end{equation}
($\alpha=\alpha_0$, $\gamma=\gamma_{0\,0\,0\,0}$ etc.) and our
 self-consistency  problem is reduced to a simple one-dimensional minimization problem which can be easily 
solved. To fourth order we
 have  a mexican hat potential when we are in the mixed state ($\alpha<0$ and $\gamma >0$) and the
 minimum for the grand potential is obtained for non-zero $\Delta_0$.  Requiring 
 $\left.\partial_{\Delta_0}(\Omega_S-\Omega_N)\right|_{\tilde{\Delta}_0}=0$ gives
\begin{equation}\label{mincon8} z^3+a_1(T,H)z^2+a_2(T,H)z+a_3(T,H)=0 \end{equation} where 
$z=\tilde{\Delta}_0^2$ and $a_1=\frac{3\kappa}{4\eta}, \; a_2=\frac{\gamma}{2\eta},\; a_3=
\frac{\alpha}{4\eta}$. $\tilde{\Delta}_0$ is the value of $\Delta_0$ which minimizes 
$\Omega_S-\Omega_N$. 
Equation~(\ref{mincon8})  is a cubic equation and can be solved exactly. To fourth order we obtain
\begin{equation} \label{mincon4} \tilde{\Delta}_0^2=-\frac{\alpha(T,H)}{2\gamma(T,H)}\; 
\; \;\Omega_S-
\Omega_N=-\frac{\alpha^2(T,H)}{4\gamma(T,H)} \end{equation}
Equation~\ref{mincon8} yields $\tilde{\Delta}_0$ and therefore $\Delta({\bf r})$ and
 $\Omega_S-\Omega_N$ as a function of $H$. The value $\tilde{\Delta}_0$ which minimizes
 $\Omega_S-\Omega_N$ will be a function of $H$ and $T$  through the coefficients 
 $\alpha, \gamma, \kappa, \eta$. Because magnetic quantization has been accounted for exactly, all
 coefficients and, hence, $\tilde{ \Delta}_0$ and  
 $\Omega_S-\Omega_N$ are oscillating functions of $H$ for a given temperature $T$.
 The condensation energy $\Omega_S-\Omega_N$ oscillates 
$180^{\circ}$ out of phase with the normal state $\Omega_N$ close to $H_{c2}$. This is the origin of the 
damping of the magnetic oscillations of $\Omega_S$ compared to $\Omega_N$. The physical reason for 
this effect is rather simple as will be explained in Sec.\ \ref{physical}. 

We now consider the magnetization $M_S \equiv\left( \partial_H \Omega_S\right)_{\mu}=
\left(\partial_H(\Omega_N+[\Omega_S-\Omega_N])\right)_{\mu}$.
 The grand potential for a free 2D electron gas $\Omega_N$  can be calculated 
analytically for the case when only two
Landau levels are partially occupied.~\cite{nor1,nor2} For relatively high $T$, low $H$ or small $g$-factor, this
 assumption breaks down but it is then straightforward  to calculate $\Omega_N$ numerically. It should be 
noted that the chemical potential $\mu$ in general is a function of $H$. We have in most of this article, for 
simplicity, kept the  chemical potential $\mu$ fixed thereby avoiding having to determine 
 $\mu$ self-consistently. The oscillatory effect of the chemical potential 
is most important for low temperatures ($T\,\raisebox{-0.4ex}{$\stackrel {<}{\sim}$}\,0.2$)
 and very clean samples 
such that higher harmonics contribute to the magnetic oscillations. In section  \ref{fixed N}
 we will show that even in this case one can to a good approximation consider the chemical potential constant 
in the mixed state.

\section{Comparison between numerical data and perturbation expansion} \label{numerical}
Recently it has been claimed that the degeneracy of the Landau levels should give rise to non-perturbative 
terms in the expression for $\Omega_S-\Omega_N$ making the Gor'kov theory invalid. For finite 
temperature there should be a non-perturbative $\Delta_0 ^3$-term in Eq.(\ref{Landau form})
 resulting in
  many interesting thermodynamic effects.~\cite{Bahcall} It is therefore of importance to establish the 
validity of the perturbation theory developed in the preceding sections so that we can use it to derive 
results instead of a cumbersome numerical solution. This is essential in the case when many 
Landau levels participate in the pairing since the computation time is very long in this regime 
for the numerical solution. 
In order to estimate  the accuracy of our perturbation expansion, we compare it to an exact  numerical 
solution of the corresponding BdG-equations. As mentioned earlier, we have set up a code which solves these 
equations self-consistently. We have chosen parameters such that $\omega_D/\omega_c=5$, 
 $\frac{V}{\hbar\omega_cl^2}=8.2$ and $k_bT/\hbar\omega_c=0.28$ when $n_f=12$. In Fig.\ 1 
we show the order-parameter $\tilde{\Delta}_0$ as a function of the magnetic field. The chemical potential 
$\mu$ is fixed. We have plotted both the numerical, the fourth 
order and the eighth order solutions. There is good agreement between the numerical solution and our 
perturbation expansion for both fourth and eighth order. The general behaviour
 of $\Delta_0$ is correctly predicted by both the fourth order and the 
eighth order expansions. In Fig.\ 2  we have plotted the condensation energy $\Omega_S-\Omega_N$.
 We are measuring energies in 
units of $\hbar \omega_c$.  It is apparent that the contribution $\Omega_S-\Omega_N$ has local minima
 for $n_f$ integer. Since  $\Omega_N$ has local maxima for $n_f$ integer the condensation energy 
 oscillates $180^{\circ}$ out of phase with the contribution from the normal state 
$\Omega_N$. We therefore get partial cancellation of the normal state oscillations and a damping of the 
dHvA-oscillations. This is seen in Fig.\ 3 where we have plotted the magnetization 
$M\equiv -\left( \partial_H \Omega \right) _{\mu}$ for both the normal state and the mixed 
state. When the 
superconducting order starts to increase at $n_f \simeq 10$, we get significant damping of the dHvA 
oscillations. Again the 
agreement with the numerical data is good as long as $n_f\,\raisebox{-0.4ex}{$\stackrel {<}{\sim}$}\,12$. 
Eighth order theory tends to agree better with numerical data than does the fourth order theory indicating 
that the perturbation expression is valid. Once we go
 too far into the superconducting state, the perturbation theory starts to disagree with the numerical
 results, also as expected. We see from Fig.\ 1  and Fig.\ 2 that the magnitude of $\Delta_0$
 and $\Omega_S-\Omega_N$ is still fairly well described for $n_f>12$, but both fourth and the eighth
 order expansions start to pick up spurious oscillations in the order-parameter and in  the energy. 
 $\Omega_S-\Omega_N$ actually starts to oscillate in phase with $\Omega_N$ 
according to the perturbation theory. This gives enhancement of the dHvA-oscillations in the mixed 
state as compared to the normal state, as  seen from Fig.\ 3. This is an unphysical effect and is absent in the 
exact solution. Since this enhancement is neither confirmed numerically nor experimentally, we conclude 
that perturbation theory in the single parameter $\Delta_0$ breaks down at this point. It can be shown~\cite
{Bruun} that the Gor'kov expansion is convergent if the change in the quasiparticle energies 
 $|E^{\eta}_{{\mathbf{k}}}-\xi_{\eta}|$  is not larger than $O(k_bT)$. We have 
looked at the numerically calculated quasiparticle energies as a function of $n_f$. As expected and in
 agreement with Norman \textit{et al.}~\cite{Big Mac1} we observe that the quasiparticle bands go from being
 essentially broadened 
Landau levels close to the transition point to loosing all their Landau level structure deeper into the mixed 
state. For the above specific case we have found that  for $n_f\:\raisebox{-0.4ex}{$\stackrel {>}{\sim}$}\:12$ 
the quasiparticle energies are changed so much  that the above condition for the validity of the Gor'kov series does not hold in large regions of ${\mathbf{k}}$-space thus 
explaining the breakdown of perturbation theory. 
We have compared the numerical solution and the perturbation expansion for a number of different 
parameters.  Our conclusion is that both  fourth and eighth  order perturbation theories describe well the 
superconducting state and the corresponding damping of the magnetic oscillations near the transition point. 
However, the perturbation theory eventually breaks down when the quasiparticle levels are changed too much, 
in the sense described above. 
The convergence range of the Gor'kov expansion is determined by the temperature $k_bT$. 

 The numerical results show total suppression of the dHvA effect once we are 
deep enough into the mixed state. In Fig.\ 3 the numerical solution shows that $M_s$ loses its 
dHvA structure completely for 
$n_f\:\raisebox{-0.4ex}{$\stackrel {>}{\sim}$}\: 12$. This contradicts the recent predictions 
of a sign shift of the first harmonic of the 
dHvA oscillations.~\cite{Maniv4} This prediction is partly based on the assumption that the quasiparticle 
spectrum can be described by a simple splitting of the Landau levels into two levels symmetrically 
placed on around each Landau level even when the actual change in energy is rather large 
 $(|E^{\eta}_{{\mathbf{k}}}-\xi_{\eta}|/\hbar\omega_c \approx \pm 0.22)$. We have found that the low lying 
quasiparticle levels loose their Landau level structure and describe essentially localized bound states 
when the change in energies is of the above magnitude. This crossover to localized states makes the argument 
leading to the sign change of the first harmonic invalid and it leads to  the  suppression 
of  the magnetic oscillations.~\cite{Big Mac3} 

\section{Damping of the magnetic oscillations}\label{damping}

\subsection{Physical interpretation}\label{physical}
To get a physical understanding of the superconducting damping of the magnetic oscillations, it is helpful to 
consider the ground-state which gives the dominant contribution to the grand potential for low temperatures.
 By analogy with the case of no magnetic field,~\cite{Schrieffer} our numerical solution is based on the
 following canonical transformation: \begin{equation} 
 \hat{\gamma}_{{\bf k} \uparrow}^{\eta}=\sum_N \left[ {u^{\eta}}^*_{N {\bf k}}\hat{a}_{N{\bf k}
\uparrow}+{v^{\eta}}^*_{N {\bf k}}\hat{a}_{N-{\bf k}\downarrow} ^{\dagger} \right] \end{equation} 
\begin{equation}  \hat{\gamma}_{{\bf k} \downarrow}^{\eta}= \sum_N \left[ {u^{\eta}}^*_{N {\bf k}} 
\hat{a}_{N{\bf k}\downarrow}-{v^{\eta}}^*_{N {\bf k}}\hat{a}_{N-{\bf k}\uparrow} ^{\dagger} \right]
\end{equation} 
where $u^{\eta}_{N {\bf k}}$ is the coefficient of $\phi({\mathbf{r}})_{N{\mathbf{k}}}$ and 
$v^{\eta}_{N {\bf k}}$ is the coefficient of $\phi^*({\mathbf{r}})_{N{\mathbf{-k}}}$ in the 
Bogoliubov amplitudes $u({\mathbf{r}})$ and $v({\mathbf{r}})$ for the $\eta$'th solution 
respectively.
The corresponding ground state of our mean field Hamiltonian is then
\begin{equation}\label{gnd} |\Psi _g\!>\propto \prod _{\eta {\bf k}} \hat{\gamma}_{{\bf k} \uparrow}^
{\eta}\hat{\gamma}_{-{\bf k} \downarrow}^{\eta}|\Psi\!> \end{equation}
 where $|\Psi\!>$ is a state with all single particle states with energy less that $\mu-\omega_D$ empty and
 all  single particle states with energy higher than $\mu+\omega_D$ occupied . We see that 
Eq.~(\ref{gnd}) gives a coherent superposition of states
 where the pairs $\hat{a}_{N {\bf k}}^{\dagger}\hat{a}_{N' -{\bf k}}^{\dagger}|0\!>$ are either occupied
 or unoccupied. When we have a Landau level at the chemical potential $\mu $ ($n_f=$integer) it does not 
cost any kinetic energy to make a superposition of states with either occupied or unoccupied pairs formed by 
electrons in that level. The instability towards superconductivity is therefore largest when we have 
$\mu=(n+1/2)\hbar \omega$. Since the grand potential of the normal state is at a 
maximum~\cite{Schoenberg} when $\mu=(n+1/2)\hbar \omega$ we have that $\Omega_S-\Omega_N$ 
and $\Omega_N$ oscillate $180^{\circ}$ out of phase. This analysis is true for both constant chemical 
potential and constant number of particles. In the latter case one works with the Helmholtz free energy but 
the conclusions are the same. Mathematically the maximum in the damping comes from the fact that when 
the chemical potential is at a Landau level the sum in equation~(\ref{alpha}) is  dominated by the 
terms with zero denominators, as an application of l' Hopital's rule on these terms confirms.
Hence $\alpha(H)$ has a local minimum and the superconducting order a local maximum. 
 This is the physical picture of the damping of the magnetic oscillations that naturally 
emerges from our formalism. 

Norman \textit{et al}~\cite{Big Mac1} interpret the damping of the 
magnetic 
oscillations as an effect of  the broadening of the Landau levels due to superconducting order. An alternate explanation has been put forward
P.\ Miller and B.\ L. Gy\"{o}rffy.~\cite{Miller} that emphasizes the role of
non--diagonal pairing. There is in fact an intimate link between the two
approaches that we now elucidate by the following simple calculation:
We estimate $\Omega_{\rm S} - \Omega_{\rm N}$ 
(for simplicity we consider $T = 0$) for the two cases
when (I) the chemical potential is at a Landau level ($n_f$ integer;
maximum of the free energy) and (II) when it is exactly between
two LL ($n_f$ is half an odd integer; minimum of the free energy).
In both cases we diagonalize the BdG equations approximately, but 
insist on using degenerate perturbation theory, because the diagonal approximation breaks down. When $n_f$ is an integer, the lowest lying 
quasi--particle excitations have the orbital character of the $n_f$--st LL,
and perturbation theory yields for the quasi--particle energy
$E_{n_f \vec k} = \vert F_{n_f \vec k}\vert$. It can easily be seen that 
the contributions of the other LL to the ground state energy  
cancel pairwise within degenerate perturbation theory (essentially because
within degenerate perturbation theory level repulsion is symmetric 
with respect to the unperturbed degenerate level). Therefore the reduction
in the maximum of the free energy   
for case (I) in the mixed state is
$$%
\Omega^I_{\rm S} - \Omega^I_{\rm N} \sim
-{1\over 2} \sum_{\vec k}\vert F_{n_f \vec k}\vert  
$$%
to lowest order in the pairing self energy.

A similar calculation for case (II),  when $n_f$ is half an odd integer
and the free energy is a minimum, gives  an energy shift
which is of higher than linear order in the pairing
self energy, because degenerate perturbation
theory now leads to complete pairwise cancelling for all Landau levels
to first order in the pairing self--energy).
Therefore
the minimum of the oscillation is reduced by substantially less
than the maximum, which shows that the damping of the oscillations
is a direct consequence of the broadening of the  quasi-particle
levels accompanied by the mixed orbital character of quasi--particle 
excitations.

\subsection{Finite Zeeman splitting}
Inclusion of spin in general reduces the 
magnitude of oscillations of $\Delta_0$ and $\Omega_S-\Omega_N$. This reduction in the amplitude of the 
oscillations is due to the fact that spin up and spin down electrons now have different energy (unless 
$g=2nm_0/m^*$  $n=0,1,2,\ldots$). We can never  have the situation whereby pairing occurs without a cost in 
kinetic energy. The oscillatory effect is therefore damped. The mathematical reason for the reduction in
 oscillations is that for finite spin the numerator in Eq.(~\ref{alpha}) never becomes  zero. So we 
expect the magnetic oscillations in the mixed state to be reduced due to spin. The question is whether this 
reduction is larger or smaller than the corresponding reduction in the normal state thus giving rise to extra 
damping effects. For temperatures (or impurity concentrations) such that only the first harmonic of the 
dHvA oscillations is important 
in both the mixed and the normal state ($k_bT\:\raisebox{-0.4ex}{$\stackrel {>}{\sim}$}
 \:0.2\hbar\omega_c$) and within the region of validity of the perturbation expansion of 
$\Omega_S-\Omega_N$ the result is that the amplitude of the first harmonic of the 
dHvA oscillations in the mixed state  is reduced by a factor $\cos(\pi\frac{gm^*}{2m_0})$. This is the same 
reduction as in the normal state and hence the relative damping due to superconductivity is insensitive to 
spin splitting. This result will be proved in section~\ref{Simple}. We 
have confirmed this result by solving the BdG-equations numerically with and without a finite 
Zeeman splitting. The reduction in the amplitude  in both the mixed and in the normal state as compared to
 the amplitude with no spin splitting  corresponds very well to a $\cos(\pi\frac{gm^*}{2m_0})$ factor in the 
region where the mixed state is described well by the perturbation expansion. Deeper into the mixed state 
the numerical results indicate that the effect of spin is  suppressed by the superconducting order. The 
reduction in the amplitude of 
the magnetic oscillations due to a finite Zeeman term is less than the $\cos(\pi\frac{gm^*}{2m_0})$ factor. 
This is due to the fact that when the superconducting order increases, the pairing interaction starts to dominate the 
Zeeman term and the effect of any finite $g$-factor is suppressed.

 So we conclude that within the region described well by our perturbative expansion a finite Zeeman term 
does not alter the rate of the damping of the magnetic oscillations due to superconductivity. When only the 
first harmonic is important the effect of the Zeeman term is simply a reduction by a factor 
$\cos(\pi\frac{gm^*}{2m_0})$ for the amplitude of the oscillations in both the mixed and in
 the normal state. Deeper into the mixed state the superconducting order starts to suppress the effect of the 
spin splitting and the magnetic oscillations is less affected by a finite Zeeman term. Hence in this region the 
relative size of the magnetic oscillations in the mixed state as compared to the normal state is larger 
for finite spin splitting and the damping is less efficient as compared to the $g=0$ case. 

\subsection{Conserved number of particles} \label{fixed N}
For two-dimensional systems with a fixed number of particles it is well known~\cite{Schoenberg} that the 
magnetic field dependence of the chemical potential $\mu(H)$ has a strong 
effect on the magnetic oscillations in a normal metal when higher harmonics are important.  
For low temperatures and clean samples the shape of the oscillations look 
qualitatively different when the chemical potential is fixed as compared to  when the 
number of particles is fixed. We have up till now mainly considered the case 
of a constant chemical potential. When the number of particles is held fixed 
we need to consider Helmholz free energy $F=\Omega+N\mu$. The chemical 
potential is determined by the equation
\begin{equation}\label{condition} <\hat{N}>=\sum_{\sigma N
{\mathbf{k}}\eta}[|u^{\eta}_{
{\mathbf{Nk}}}|^2f_{\sigma {\mathbf{k}}}^{\eta}+|v_{ {\mathbf{Nk}}}^{\eta}
|^2(1-f_{-\sigma {\mathbf{k}}})]=N \end{equation}
where $f^{\eta}_{\sigma {\mathbf{k}}}=(\exp(\beta E^{\eta}_{\sigma {\mathbf{k}}})+1)^{-1}$ 
This is a numerically cumbersome problem since we need to solve the BdG-
equations self-consistently for a given chemical potential, then calculate 
$<\hat{N}>$ and repeat the calculation for a new value of $\mu$ until 
Eq.(\ref{condition}) is obeyed. However it is essential that we determine 
the chemical potential self-consistently. If we naively assume that the 
chemical potential oscillates as in the normal state we would obtain persistent 
magnetic oscillations of the free energy even when the Landau level structure 
is completely destroyed by superconducting order. In Fig.\ 4 we 
have plotted the magnetization when the chemical potential is constant ($\Box$) and when the number
 of particles is constant ($\ast$) for a very 
low temperature. We have chosen parameters such that 
 $\omega_D/\omega_c=5$,  $\frac{V}{\hbar\omega_cl^2}=9.0$ and 
 $k_bT/\hbar\omega_c=0.05$ and $gm^*/m_0=1$ when $n_f=12$. For comparison 
the solid and dotted lines give the magnetization in the normal state for 
 $n_f\:\raisebox{-0.4ex}{$\stackrel {>}{\sim}$}\:8.2$ for conserved $\mu$ and $N$ 
respectively. We see that there is only a 
significant difference between the two curves close to $H_{c2}$ ($n_f\approx 7.7$ at 
$H_{c2}$) 
when the chemical potential behaves differently in the two cases. Deeper into the 
mixed state the oscillatory behaviour 
of the chemical potential is damped by the superconducting order and it 
becomes practically constant. This is illustrated in Fig.\  
5 where we have plotted $n_f=\mu(H)/\hbar\omega_c-0.5$ as a function of the 
magnetic field ($H_{c2} \approx 1.5$) when 
the number of particles $N$ is constant (solid line) and when the chemical 
potential is constant (dashed line). We see that the oscillations in the chemical potential 
when $N$ is constant are damped in the mixed state. Once the superconducting 
order has damped the oscillations in the magnetization it has also damped 
the oscillation in $\mu(H)$ and the behaviour for fixed N is essentially the same 
as for fixed $\mu$. Thus the conclusion is that although there is 
some difference in the dHvA signal close to $H_{c2}$ when $N$ is fixed conserved as 
opposed to fixed $\mu$ the overall rate of damping of 
the oscillations is the same in the two cases.

\section{Simplified form for the damping} \label{Simple}
\subsection{The first harmonic of the condensation energy} \label{simple}
To obtain a simple form for the damping, we must take a closer look at the coefficients $\alpha(H)$
 and $\gamma(H)$ given in Eq.\ (\ref{alpha}) and Eq.\  (\ref{gamma}). As mentioned already, the 
transition to the mixed state occurs when $\alpha(H)$ 
changes sign. The Gor'kov expansion is most relevant for temperatures such that only the 
lowest harmonics 
of the dHvA signal are significant. This allows us to focus only on the zeroth and first harmonics of the relevant 
quantities. Thus we take $\alpha(H)$ to have the form:
\begin{equation} \label{alphaa}\alpha(H) \simeq a_1 (1-H_{c2}/H)+a_2 
\cos (2\pi \mu/\hbar \omega_c ) \end{equation}
where $a_1 >0$ and $a_2>0$. The coefficients $a_1$ and $a_2$ 
will in general depend weakly on the magnetic field but we assume they are constant. This is reasonable since
 for $\mu/\hbar \omega_c \gg 1$ the rate of change of $a_1$ and $a_2$ is very slow as  compared to 
the frequency $\mu mc/\hbar e$ of the oscillations. The essential physics comes from the sign change of 
$\alpha(H)$ and 
its oscillatory behaviour, combined with the features of $\gamma(H)$ described below. For simplicity 
we confine ourselves to fourth-order perturbation theory. The fourth-order coefficient  $\gamma(H)$ is 
has  the  form:
\begin{equation} \label{gammaa} \gamma(H) \simeq g_1-g_2\cos (2\pi \mu/\hbar \omega_c ) 
\end{equation} 
where $g_1>0$ and $g_2>0$.
Again both $g_1$ and $g_2$ depend on the magnetic field but this dependence is weak as  compared 
to the strong oscillatory behaviour coming from the Landau level structure. Note that we have opposite
signs for the first harmonics of  $\alpha(H)$ and $\gamma(H)$.
  In section~\ref{a2g2} we will extract estimates of $a_2$ and $g_2$ from Eq.~(\ref{alpha}) and 
Eq.~(\ref{gamma}) whereas the approximate expressions for $a_1$ and $g_1$ will be given in appendix
 \ref{a1g1}.
Using these approximate forms for $\alpha(H)$ and $\gamma(H)$ we get for the condensation energy
\begin{equation} \Omega_S-\Omega_N=-\frac{\alpha^2(T,H)}{4\gamma(T,H)} \simeq 
-\frac{(a_1 (1-H_{c2}/H)+a_2 \cos (2\pi \mu/\hbar \omega_c ))^2}{4(g_1-
g_2\cos (2\pi \mu/\hbar \omega_c ))}. \end{equation}
Assuming that $g_2 \ll g_1$ we get the following approximate form for the first harmonic of
$\Omega_S-\Omega_N$ to first order in $g_2/g_1$:
\begin{equation} \label{1harms} \left(\Omega_S-\Omega_N \right)_{1} \simeq \frac{1}{4}
\left[\frac{2a_1a_2}{g_1}(H_{c2}/H-1)-\frac{g_2a_1^2}{g_1^2}(H_{c2}/H-1)^2 
-\frac{3g_2a_2^2}{4g_1^2} \right] \cos (2\pi \mu/\hbar \omega_c) \end{equation}
where $\Omega(H)_n$ is the n'th harmonic of $\Omega(H)$. It should be recalled that the above expression
 is only valid for $\alpha(H) <0 $. When we are deep enough into the superconducting state so that we are
 away from the reentrance region we have 
 $a_1(H_{c2}/H-1)>a_2$.  This means $\frac{3a_2^2g_2}{4g_1^2}\ll\frac{a_2^2}{g_1}<
\frac{a_1a_2}{g_1}(H_{c2}/H-1)$ and we can neglect the small constant term 
$\frac{3a_2^2g_2}{g_1^2}$. We thus get the following form for the first harmonic of the grand potential:
\begin{equation} \label{1hg}{\Omega_S}_1={\Omega_N}_1+
\left(\Omega_S-\Omega_N\right)_{1} \end{equation}
where~\cite{Schoenberg}
 \begin{equation}\label{1harmn} \Omega_{N_1}=-\frac{eL_xL_yH}{2\pi\hbar c}
\frac{\hbar \omega_c}{\pi^2}\left[4\pi^2\frac{k_bT}{\hbar \omega_c}e^{-2\pi^2\frac{k_bT}{\hbar
\omega_c}}\right]\cos (2\pi \mu/\hbar \omega_c )\end{equation}
We have written the reduction due to finite temperature in square brackets.

\subsection{Calculation of the oscillatory terms}\label{a2g2}

In this section we will derive some approximate expressions for the 
coefficients $a_2$ and $g_2$. We are interested in how $a_2$ and $g_2$ depend on the parameters $n_f$, 
 $\omega_D$, and  $T$. It turns out that it is fairly straightforward to extract this dependence for the
 oscillatory terms. First we note the following approximate identity coming from the law of large numbers:
\begin{equation} \label{Bapprox} B_0^{n_1,n_2} \simeq \frac{1}{\sqrt[4]{\pi n_1}}
e^{-(n_1-n_2)^2/8n_1} \simeq\frac{1}{\sqrt[4]{\pi n_f}}
e^{-(n_1-n_2)^2/8n_f} \end{equation}
where we have assumed that $|n_1-n_2|/n_1 
\ll 1$ and $n_1 \simeq n_f$ (i.e $\omega_D/\omega_c\ \equiv 2\sigma \ll n_f$). 
Using this formula, Eq.~\ref{alpha}, and the Poisson identity we get the following integrals for $\alpha(H)$:
\begin{eqnarray}\label{Ill} \hbar \omega_c \sum_{n_1,n_2}{B_0^{n_1 \:n_2}}^2
\frac{\tanh(\beta\xi_{n_1}/2)+\tanh(\beta\xi_{n_2}/2)}{\xi_{n_1}+\xi_{n_2}}w^2(n_1)w^2(n_2)
=\nonumber\\
\sum_{l,m} e^{2\pi i n_f(m-l)}\int dx \int dy e^{2\pi i(mx-ly)}
\frac{e^{-\frac{(x-y)^2}{4n_f}}}{\sqrt{\pi n_f}}\frac{\tanh(\beta\xi_x/2)
+\tanh(\beta\xi_y/2)}{x+y} w^2(x)w^2(y)=\nonumber \\
\sum_{l,m} I_{l,m}e^{2\pi i n_f(m-l)}\end{eqnarray}
where $\xi_x=\hbar\omega_cx$ and $w(x)=e^{-(\xi_x/0.5\hbar\omega_D)^2}=e^{-x^2/\sigma^2}$.
To estimate $I_{l,m}$ where $(l,m) \neq (0,0)$ we write the integral in the form:
\begin{equation} I_{l,m}=\frac{2k_bT}{\hbar\omega_c}\sum_{\nu}\int dx \int dy
\frac{e^{-(x-y)^2/4n_f}}{\sqrt{\pi n_f}} \frac{e^{-2(x^2+y^2)/\sigma^2+2\pi i
(mx-ly)}}{(x-i\omega'_{\nu})(y+i\omega'_{\nu})} \end{equation}
where $\omega'_{\nu}=\omega_{\nu}/\omega_c$. The first harmonic of  $\alpha(H)$
 comes from the terms with $|l-m|=1$. Taking $m=1$ and $l=0$ yields the integral:
\begin{equation} \int dx \frac{e^{2\pi ix-2x^2/\sigma^2}}{x-i\omega'_{\nu}}
\int dy \frac{e^{-2y^2/\sigma^2-(x-y)^2/4n_f}}{y+i\omega'_{\nu}}\end{equation}
We approximate this integral by:
\begin{equation} \int dx \frac{e^{2\pi ix}}{x-i\omega'_{\nu}}\int dy
\frac{e^{-(y-x)^2/4n_f}}{y+i\omega'_{\nu}} \end{equation} since we have assumed   
$8n_f \ll \sigma^2$. The integral can be solved exactly and we get :
\begin{eqnarray} I_{0,1}&=&\frac{4k_bT\pi^2}{\hbar\omega_c\sqrt{\pi n_F}}
\sum_{\nu\geq 0} e^{-2\pi\omega'_{\nu}}\left([1-\Phi(\omega'_{\nu}/\sqrt{n_f})]
e^{\omega'^2 _{\nu}  /n_f}+e^{t^2}[(1-\Phi(t))]e^{-4\pi^2 n_f}\right)\ \nonumber \\ 
&\simeq& \frac{4k_bT}{\hbar\omega_c\sqrt{\pi n_f}}\pi^2 
e^{-2\pi^2\frac{k_bT}{\hbar\omega_c}}  \end{eqnarray}  where 
 $\Phi(x)\equiv 2\pi^{-1/2}\int_0^xe^{-s^2}ds$ is the error function  and $t=(\omega'+2\pi n_f)/\sqrt{n_f}$.  
Here we have used that $\exp (-2\pi\omega_{\nu\neq 0})\ll \exp (-2\pi\omega_{\nu= 0})$ for 
 $2\pi^2k_bT/\hbar \omega_c \:\raisebox{-0.4ex}{$\stackrel {>}{\sim}$}\: 1$, $\exp(-4\pi^2n_f)\ll 1$, and 
 $n_f^{-1/2}\omega_{\nu=0} \ll 1$.  So, in this temperature 
range the dominant contribution to the first harmonic comes from the lowest Matsubara frequency, 
which makes our approximation above self-consistent. The contribution to
the first harmonic from the $|l-m|=1$ term given $ l, m\neq 0$ can be calculated in the same way; it 
is proportional to $\exp(-2\pi^2 m kT/\hbar\omega_c)$ and therefore negligible
for $2\pi^2k_bT/\hbar \omega_c \:\raisebox{-0.4ex}{$\stackrel {>}{\sim}$}\: 1$ 
in agreement with the results obtained by 
Gruenberg \textit{et al}.~\cite{Gruenberg}After some algebra, the 
calculations outlined above combined with Eq.~(\ref{alpha}) lead to 
the following  approximate result:
\begin{equation}\label{a_2}
a_2 \simeq \frac{V^2\left( \frac{a_x}{l} \right)2\pi}{L_x L_yl^2\sqrt{\pi n_f}}
\frac{k_bT}{(\hbar \omega_c)^2}
e^{-2\pi^2\frac{k_bT}{\hbar \omega_c}} \end{equation}
The above result that  $a_2$ is proportional to $1/\sqrt{n_f}$ and 
 $k_bT e^{-2\pi^2\frac{k_bT}{\hbar \omega_c}}$ and independent on $\omega_D$
 is still correct even when $\sigma^2\ll 8n_f$,  as long as $\min(\sigma,\sqrt{n_f})\gg 1$ and 
$2\pi^2k_bT/\hbar \omega_c \:\raisebox{-0.4ex}{$\stackrel {>}{\sim}$}\: 1$. 
 Inclusion of spin is equivalent to making the substitution
 $x \rightarrow x+\frac{gm^*} {4m_0}$ and $y \rightarrow y-\frac{gm^*}{4m_0}$ in the integrals 
 $I_{l,m}$. This results in a reduction factor $\cos(\pi \frac{gm^*}{2m_0})$ in Eq.(~\ref{a_2}) if 
$\min(\sqrt{n_f}, \sigma) \gg g$.

 The calculations for $g_2$ are very similar to the ones above. Using Eq.~(\ref{gamma}) and the Poisson
 formula we end up with the following integrals determining the dependence of $\gamma$ on $n_f$, $T$ 
and $\omega_D$:
\begin{eqnarray} \label{gint} \sum_{\stackrel{l_1,l_2,l_3,l_4}{\nu}}\frac{k_bT}{n_f}
e^{2\pi in_f(l_1+l_3-l_2-l_4)}
 \int dx_1\!\ldots dx_4 \frac{e^{[(x_1-x_4)^2+(x_3-x_2)^2+(x_1-x_2)^2+
(x_3-x_4)^2]/8n_f}}{(i\omega'_{\nu}-x_1)(i\omega'_{\nu}+x_2)(i\omega'_{\nu}-x_3)
(i\omega'_{\nu}+x_4)}\times 
\nonumber\\  e^{-2(x_1^2+x_2^2+x_3^2+x_4^2)/\sigma^2}e^{2\pi i (l_1x_1+l_3x_3-l_2x_2-l_4x_4)}
\Xi^{x_1+x_4,x_2+x_3}_{x_1+x_2,x_3+x_4} 
\end{eqnarray}
 where $\Xi^{j_1,j_2}_{j_3,j_4}$ is given in Eq.~(\ref{xi}).
Contributions to the first harmonic $g_2$ come from the terms with $|l_1+l_3-l_2-l_4|=1$. As in the 
case for $a_2$ we can  neglect the terms with  more than one $l_i$ different from zero when 
 $2\pi^2k_bT/\hbar \omega_c \:\raisebox{-0.4ex}{$\stackrel {>}{\sim}$}\: 1$. Although we do not have any
 simple expression for $\Xi^{x_1+x_4,x_2+x_3}_{x_1+x_2,x_3+x_4}$ we can
 still extract the dependence on $T$, $\omega_D$, and $n_f$. This is because  the integral over 
 $x_2\ldots x_4$ does not vary appreciably  with $x_1$ on a scale $\simeq \omega'_{\nu=0}$.  Using the 
  result $\int dx \frac{exp(2\pi ix_1)}{i\omega'-x_1}f(x_1) \propto f(0)e^{-2\pi \omega'}$ ($\omega' >0$) 
for any well-behaved function $f(x)$ which varies slowly for $x\,\raisebox{-0.4ex}
 {$\stackrel {<}{\sim}$}\,\omega'$ and taking $l_1=1,\ l_2=l_3=l_4=0$ we get the integral:
\begin{eqnarray} \frac{k_bT}{n_f}\int dx_1\frac{e^{2\pi ix_1}}{i\omega'_{\nu}-x_1}\int dx_2\ldots dx_4
\frac{e^{[x_4^2+(x_3-x_2)^2+x_2^2+(x_3-x_4)^2]/8n_f}}{(i\omega'_{\nu}+x_2)(i\omega'_{\nu}-x_3)
(i\omega'_{\nu}+x_4)} \nonumber \\ e^{2(x_2^2+x_3^2+x_4^2)/\sigma^2}
\Xi^{x_4,x_2+x_3}_{x_2,x_3+x_4} 
\end{eqnarray} 
The factors   $1/(i\omega' \pm x_j)$ in the integrand makes the integral largely independent of any long 
range behaviour determined by $\sigma$ and $n_f$ as long as $|\omega'| \ll \min(\sigma,  \sqrt{n_f})$. 
 We therefore conclude that $g_2$ is independent of $\omega_D$ and that it 
only depends on  $n_f$ through the $n_f^{-1}$-factor coming from the four $B_0^{N,M}$ coefficients.
 We also obtain that $g_2$ is proportional to $k_bT \exp (-2\pi^2\frac{k_bT}{\hbar \omega_c})$. 
The proportionality constant is found through an exact evaluation of $\gamma$ given in 
Eq.~(\ref{gamma}). We obtain:
\begin{equation} \label{g_2} g_2 \simeq \frac{\left(\frac{V}{\hbar\omega_c}\right)^4 27}
{n_f (L_xL_y)^3 l^2}
k_bTe^{-2\pi^2 \frac{k_bT}{\hbar \omega_c}} \end{equation}
Again the effect of spin (ie.\ non-zero $g$-factor) provide an additional  $\cos(\pi\frac{gm^*}{2m_0})$ in 
Eq.(~\ref{g_2}). It is not surprising that the oscillatory terms $a_2$ and $g_2$ are independent of the 
pairing width $\omega_D$  
since the oscillations are a consequence of the individual Landau levels going through the chemical potential.
 Likewise the $1/\sqrt{n_f}$ and $1/n_f$ dependence reflect the fact that the probability for two electrons, 
each with  energy $(n+1/2)\hbar \omega_c$, to 
form a pair with minimum center-of-mass energy is proportional to $1/\sqrt{n}$ for high quantum
 numbers, as can be seen from Eq.(\ref{Bapprox}). This proportionality can be explained via simple phase-space
 considerations. 
We have tested the dependence of $a_2$ and $g_2$ on the different parameters $n_f$, $\omega_D$ and 
$T$ and we find excellent agreement with our approximate forms.

To facilitate comparison with earlier papers we will now formally treat the order parameter 
$\Delta({\mathbf{r}})$ as a free parameter and assume that the oscillatory behaviour of 
Eq.(\ref{Landau form}) only comes from the harmonics of the expansion coefficients 
$\alpha,\ \gamma$ etc.
This is of course incorrect since the self-consistent order parameter itself is a oscillatory function of the 
field, making the results where the  corrections to the harmonics of the dHvA oscillations due to 
superconductivity are expressed as a power series in $\Delta$~{\cite{Maki,Stephen,Maniv3} of limited 
validity. However, to compare with the earlier predictions we ignore for the moment the oscillations 
in  $\Delta_0$ and treat it formally as a free paramter (i.e.
$ (\Omega_S-\Omega_N)_1=a_2\Delta_0^2-g_2\Delta_0^4+\ldots $). Here we focus on the 
$\Delta^4$-term since there are discrepancies between the predictions of different authors for 
this term. Since 
\begin{equation}
 \Delta^2\equiv (L_xL_y)^{-1}\int d{\mathbf{r}}|
\Delta({\mathbf{r}})|^2 =\frac{V^2a_x}{\sqrt{2}L_x^2L_y^2l}\Delta_0^2 \end{equation}
  we obtain using Eq.(\ref{g_2}) and Eq.(\ref{1harmn}) the formal result for the fourth order term:
\begin{equation}\left.(\Omega_S-\Omega_N)_1\right|_{\Delta^4-term} =
-g_2\Delta_0^4\approx {\Omega_N}_1 \frac{10}{n_f}\left( \frac{\Delta}{\hbar 
\omega_c}\right)^4 
\end{equation}
Stephen~{\cite{Stephen}} obtained $\sim 16\Omega_{N_1}/n_f(\Delta/\hbar\omega_c)^4$ for the 
same quantity using a different semiclassical approach. The $n_f$ dependence of the two result agree 
but the numerical prefactors are somewhat different. The above arguments for the $n_f$ 
dependence of $g_2$ can 
easily be generalised yielding that the $n_f$ dependence of the first harmonic of the 
$\Delta^{2n}$-term 
is ${n_f}^{-n/2}$. This $n_f$ dependence agrees with the result obtained by Stephen whereas 
it disagrees with the $n_f^{-3/2}$-dependence for the $\Delta^4$-term  obtained by 
Maniv \textit{et al}.~\cite{Maniv3} We cannot overemphasize the fact
 that the above scheme to calculate the damping of the oscillations due to 
superconductivity is incorrect, since it ignores the oscillations in $\Delta$ as a function of the field. 
To include those we have to use a self-consistent order parameter and hence Eq.(\ref{1harms}).

One debated issue is the possibility of reentrance for type-II 
superconductors.~\cite{Tesanovic}
 The oscillatory behaviour of $\alpha$ due to the Landau level structure gives
 rise to the possibility of several solutions to $\alpha(H)=0$ for a given 
temperature. This should be reflected in a highly oscillatory behaviour of 
the transition line $H_{c2}(T,H)$. Such an oscillatory behaviour has never
 been observed experimentally. Using the approximate expressions for $a_1$ and
 $a_2$ we can estimate the temperature below which there is reentrance 
and such oscillations in $H_{c2}$ should occur in a two dimensional metal. 
We obtain that when there is 
no impurity scattering, no  Zeeman splitting, and $n_f \sim O(10^2)$ one should 
observe these oscillatory effects in $H_{c2}$ in a 2D metal for temperatures 
lower than $k_bT/\hbar\omega_c \approx 0.3$. However, inclusion of spin reduces
 the amplitude of the oscillations of $\alpha$ by a factor $\cos(\pi\frac{gm^*}{2m_0})$
 close to the transition line. 
Assuming that impurities reduces the oscillations by a factor $exp(-2\pi^2 k_b
 T_D/\hbar\omega_c)$ where $T_D$ is the Dingle temperature we obtain that 
there will not be any reentrance if $k_bT_D/\hbar\omega_c\approx 0.2$ no 
matter how low the temperature is. In the case of the experiments being done 
on   $\kappa\!-\!(ET)_2Cu(NCS)_2$~\cite{Caulfield} the experimental parameters are such that 
 $k_bT_D/\hbar\omega_c\approx 0.27$ and $|\cos(\pi\frac{gm^*}{2m_0})| \approx
0.3$. They will therefore never observe these reentrance effects. The magnetic
 oscillations in the thermodynamic quantaties will of course still be there since 
$\alpha$ and $\gamma$ are still oscillatory.

\subsection{Approximate results for damping}
\label{approx}
In this section we will draw some conclusions from the  general form of  the damping of the dHvA-oscillations 
due to the growth of the superconducting order described by Eq.~(\ref{1harms}). The first
 thing we notice is that in this approximation the superconducting damping has a simple polynomial form in 
 $(H_{c2}/H-1)$. The damping is maximum for $(H_{c2}/H-1)=\frac{a_2g_1}{a_1g_2}$. For 
 $(H_{c2}/H-1)>\frac{a_2g_1}{a_1g_2}$ the damping decreases when we go deeper into
the superconducting state and for $(H_{c2}/H-1)>\frac{2a_2g_1}{a_1g_2}$ the magnetic 
oscillations are enhanced by the superconducting order. This explains the observations made in 
section {\ref{numerical}. The in-phase oscillations between the fourth 
 order $\Omega_S-\Omega_N$ and $\Omega_N$ are due to the oscillatory behaviour of $\gamma(H)$.
Since $\gamma(H)$ oscillates in phase with $\Omega_N$ we will get the enhancement of the 
oscillations of $\Omega_S$ compared to $\Omega_N$ when the smooth part of $\alpha(H)$ is 
sufficiently large. Again we must emphasize that this  is obviously a sign that our perturbative
 scheme has broken down and does not reflect any physical effect.
 
To make any quantitative predictions we need to use our approximate expressions for $a_i$ and $g_i$.
Since we only have very good approximations  for $a_2$ and $g_2$ and for the temperature and spin 
dependence of $a_1$ and $g_1$ we will concentrate on properties that can be derived from these results. 
From Eq.~(\ref{1harms}) and the temperature dependence of $a_i$ and $g_i$ we conclude that the first 
harmonic of the condensation energy  
 $(\Omega_S-\Omega_N)_{1}$ is proportional to $ k_bT\exp (-2\pi^2\frac{k_bT}{\hbar \omega_c})$.
 Since we also have ${\Omega_N}_1\propto k_bT \exp(-2\pi^2\frac{k_bT}{\hbar \omega_c})$ this 
means that the magnetic oscillations have the same temperature dependence in the mixed state as in the 
normal state. This result agrees with the general theory (see Schoenberg~\cite{Schoenberg} Sec. 2.5 and 
 Sec. 2.3) valid for any part of the grand potential which is proportional to $\cos(\mu/\hbar\omega_c)$. It is
 also confirmed by experimental observations.~\cite{Corcoran2} Likewise the
 effect of spin on  $(\Omega_S-\Omega_N)_{1}$ is a reduction in the amplitude by a factor 
$\cos(\pi\frac{gm^*}{2m_0})$. This is the same reduction factor as for the oscillations in the normal 
state.~\cite{Schoenberg} We thus have no extra damping effects due to spin close to the transition line where 
the perturbation theory is valid.
 
We can now examine whether the arguments based on the Gor'kov expansion leading to a sign change of the 
first harmonic are valid. Naively one would expect a sign change since the 
 contribution from the condensation energy to the magnetic oscillations is in antiphase with the normal
 state oscillations.
When the system is deep enough into the mixed state the superconducting oscillations would dominate 
leading to a sign change of the magnetic oscillations. Extrapolating the rate of the damping close to $H_{c2}$ 
obtained from the Gor'kov expansion Maniv \textit{et al}~\cite{Maniv4} have estimated the magnetic field 
$H_{inv}<H_{c2}$ at which this sign change should occur. We are now able to show that this argument based 
on the perturbative expansion of the grand potential is incorrect. From Eq.~(\ref{1harms}) we obtain that the 
maximum amplitude of the antiphase oscillations of  $\Omega_S-\Omega_N$ is given by 
$\frac{a_2^2}{4g_2}$. Using our approximate expressions for $a_2$ and $g_2$ we get
\begin{equation} \frac{a_2^2}{4g_2}\simeq \frac{L_xL_ya_x^2\pi}{l^427}k_bT
e^{-2\pi^2\frac{k_bT}{\hbar\omega_c}}\end{equation}
Comparing this amplitude with the  contribution from the normal state oscillations given in 
Eq.~(\ref{1harmn}) 
we see that our perturbation scheme roughly predicts a maximum damping of 50\%.  It must be emphasized
 that this does not mean that the damping of the model described by the 
Hamiltonian in Eq.(~\ref{hamilton}) has a maximum of 50\%. 
 However,  using the result above combined with the results in section \ref{numerical}, we can conclude that 
neither the argument based on the Gor'kov expansion  nor the arguments based on a 
simplified form for the quasiparticle spectrum leading to an inversion of the first 
harmonic of the dHvA signal are valid.

\section{Comparison with experiment}
\label{comparison}
In this section we present a typical result for the damping of the magnetic oscillations obtained from our 
theory when many Landau levels participate in the pairing. We have chosen parameters such 
that we can compare our result with the experimental observations made by van der Wel 
\textit{et al}.~\cite{Caulfield} First we  compare our approximate expressions for 
the damping from Eq.(\ref{1harms}) with 
the result based on the exact evaluation of $\alpha$ and $\gamma$ from Eq.~(\ref{alpha}) and 
Eq.~(\ref{gamma}). We used the a set of parameters such that $k_bT/\hbar\omega_c=0.25$, 
$\frac{V}{\hbar\omega_c}=2.315$, and $\omega_D/\omega_c=75$ when $n_f=175$. There is 
no Zeeman effect and the chemical potential is conserved. In Fig.\ 6 we have
 plotted the 
magnetization for both the normal state and the mixed state calculated from the perturbative expansion to 
fourth order as a function of $n_f$. The perturbation theory predicts a substantial damping of the oscillations
 over many periods reaching a maximum for $n_f \simeq 170$. At the maximum the first harmonic is 
damped approximately 50 \% in agreement 
with the result in the previous section. As we go deeper into the mixed state, the damping decreases according 
to the perturbative scheme. Based on the results in Section~\ref{numerical}, we expect the perturbation 
theory to describe the damping well for $n_f \raisebox{-0.4ex}{$\stackrel {<}{\sim}$} 170$ . 
Due to the large number of Landau levels involved in pairing, we have not undertaken the exact numerical 
calculation for this set of 
parameters. In Fig.\ 7  we have plotted $M_s$ calculated from the exact evaluation of 
$\alpha(H)$ and $\gamma(H)$ and calculated from Eq.(\ref{1harms}). We see that the simplified 
expression reproduces the perturbative predictions well.

The above parameters approximate the experiment performed  by van der Wel 
\textit{et al}.~\cite{Caulfield} on the essentially
 2D organic superconductor $\kappa\!-\!(ET)_2Cu(NCS)_2$. 
To compare with the experimental data we will formulate our results in terms of a field dependent 
quasiparticle scattering rate $\tau$ defined such that $e^{-\pi/\omega_c\tau}$ gives the damping of the 
first harmonic of the dHvA oscillations due to superconductivity. From Eq.~(\ref{1hg}) we get:
\begin{equation} \label{scat}\tau^{-1}=-\frac{\omega_c}{\pi}\ln( 1+(\Omega_S-\Omega_N)_1/
{\Omega_N}_1) \simeq-\frac{\omega_ca_1a_2}{2\pi g_1{\Omega_N}_1}(H_{c2}/H-1)\end{equation}
where we have used Eq.~(\ref{1harms}). The approximate equality is only valid for 
 $\frac{a_2g_1}{a_1g_2} \ll H_{c2}/H-1$. Using the   expressions for $a_i$, $g_i$, and 
${\Omega_N}_1$ we can now compare this expression with the 
experimental observations. Unfortunately the experimental value of $H_{c2}$ is uncertain. 
 The transition from the normal state to the superconducting state occurs over a field range of 
approximately $2T$.~\cite{Dphil}  This gives a 'smooth' variation of the $\tau^{-1}$ on 
entering the mixed state which our theory cannot account for. To model this transition 
region we use the method introduced in ref. 2 by including a Gaussian spread in $H_{c2}$ 
In Fig. 8 we have 
plotted the experimental data for $\tau^{-1}$ (bars) measured in (THz) as a function of 
$1/B$ measured in Tesla$^{-1}$.  The solid line is our theoretical prediction based on 
Eq.(\ref{scat})  including a 
Gaussian spread in $H_{c2}$. The agreement between theory and experiment is good. It 
should be noted that we 
have no fitting parameters apart from $H_{c2}$. However, without a more 
reliable measurement of $H_{c2}$ a precise comparison between our theory and the 
experimental observations cannot be made. 

 \section{Conclusion} \label{conclusion}
In this paper we have examined the dHvA oscillations in the mixed state of a type II 
superconductor in the 2D limit using both a numerical solution of  the BdG equations and an analytical theory based on 
a self-consistent Gor'kov expansion. The use of translational and rotational 
symmetry has simplified the analysis such that we have been able to calculate the expansion
 coefficients exactly to any order without using semiclassical or other approximations. 
Comparison with the exact numerical solution has showed that perturbation theory works 
well close to $H_{c2}$ thereby disproving recent claims of non-perturbative effects. We 
have found that the condensation energy oscillates in antiphase with the normal grand 
potential, thus producing  damping of the dHvA oscillations in agreement with numerical and 
experimental results.  The damping is directly connected with the 
enhancement of superconductivity when we have a Landau level at the chemical potential. 
We have excluded  the possibility of a sign change of the first harmonic of the dHvA oscillations
 in the mixed state. The effect of spin and a conserved number of particles as opposed to a 
conserved chemical potential was examined. Using a simple  approximate form of our analytical 
theory valid when many Landau levels
 participate in pairing we have compared our theory with an experiment on the quasi 2D 
organic superconductor $\kappa\!-\!(ET)_2Cu(NCS)_2$. We have found good agreement. 
However, due to experimental uncertainty  about $H_{c2}$ any quantitative comparison is 
impossible.

\section{Acknowledgements}
We would like thank J.~Singleton and  S.~Hayden  for many helpful discussions. This work has been supported in part by EPSRC grant
GR/K 15619 (VNN nad NFJ) and by The Danish Research Academy (GMB).
\appendix

\section{} \label{68terms}
Using the symmetry of the vortex lattice and making the restriction  $\Delta_{j\neq 0}=0$ 
we obtain:
\begin{eqnarray} 
\Omega_6=-\frac{V^6a_x^6}{24(L_xL_y)^6}\Delta_0^6\sum_{n_1 \ldots n_6} f(n_1,\ldots,n_6)\nonumber \\ B_0^{n_1\,n_2}B_0^{n_3\,n_4}B_0^{n_5\,n_6}B_0^{n_5\,n_4}B_0^{n_3\,n_2}B_0^{n_1\,n_6}\Xi^{n_1+n_2,n_3+n_4,n_5+n_6}_{n_5+n_4,n_3+n_2,n_1+n_6}   
\end{eqnarray} where
\begin{equation}f(n_1,n_2,\ldots,n_{2l})=\frac{1}{\beta} \sum_{\nu} \left[ (-i\hbar \omega_{\nu}-\xi_{n_1 \downarrow})  (i\hbar \omega_{\nu}-\xi_{n_2 \uparrow})\ldots (i\hbar \omega_{\nu}-\xi_{n_{2l} \uparrow}) \right]^{-1}\end{equation} and
\begin{equation}\Xi^{j_1,\ldots j_l}_{j_{l+1}, \ldots j_{2l}}=\sum_{{\bf k} \in MBZ} \chi_{j_1}({\bf k})\cdots \chi_{j_l}({\bf k})  \chi^{\ast}_{j_{l+1}}({\bf k})\cdots \chi^{\ast}_{j_{2l}}({\bf k}) \end{equation} 
\begin{equation} \raisebox{.5 ex}{$\chi$}_j({\bf k}) =\sqrt{l}\sum_b
e^{2ik_xa_xb} e^{-i\pi b^2 /2}\phi_j (\sqrt{2}(k_y l+ba_x/l))  \end{equation}
Likewise the eighth order term gives for $\Delta_{j\neq 0}=0$:
\begin{eqnarray}\Omega_8=\frac{V^8a_x^8}{64(L_xL_y)^8}\Delta_0^8\sum_{n_1 \ldots n_8} f(n_1,\ldots,n_8)\nonumber \\ B_0^{n_1\,n_2}B_0^{n_3\,n_4} B_0^{n_5\,n_6} B_0^{n_7\,n_8} B_0^{n_7\,n_6} B_0^{n_5\,n_4}B_0^{n_3\,n_2}B_0^{n_1\,n_8} \Xi^{n_1+n_2,n_3+n_4,n_5+n_6,n_7+n_8}
_{n_7+n_6,n_5+n_4,n_3+n_2,n_1+n_8}
\end{eqnarray} The Matsubara sums and the \textbf{k}-sums can be calculated as in the fourth order case.

\section{} \label{a1g1}
In this appendix we will extract  the dependence of  $a_1$ and $g_1$ on $n_f$, $T$, $\sigma$ and spin.
 This is considerably harder than for $a_2$ and  $g_2$ because we do not have any oscillatory factor in the
 integrals that would make the long range behaviour of the remaining integrand insignificant. It turns out
 that it is still fairly straightforward to derive the temperature and spin dependence of $a_1$ and $g_1$, 
whereas we have to make some rather drastic approximations to obtain the dependence on $n_f$ and
 $\sigma$ for $g_1$. 

 The smooth part (zero harmonic) of $\alpha(H)$ comes from the terms $I_{l,l}$ in Eq.(\ref{Ill}). 
We first look at the term $l=m=0$. Making the variable substitution $v=\frac{x+y}{\sigma\sqrt{2}}$,
$u=\frac{x-y}{\sigma\sqrt{2}}$ we get the following integral:
\begin{equation} \label{smooth} I_{0,0}=\frac{\sigma}{\sqrt{2\pi n_f}} \int du\int dv 
e^{-(\frac{\sigma^2}{2n_f}+2)u^2}(\tanh[K\sigma(v+u)]+\tanh[K\sigma(v-u)])
\frac{e^{-2v^2}}{v}\end{equation}
where $K=\beta \hbar\omega_c/2\sqrt{2}\gg 1$ determines the temperature dependence of the integral.
  Since $K$ is only important around the region $v\simeq 0$ which does not contribute significantly to the 
integral, we conclude that $I_{0,0}$  is independent of the temperature to a very good approximation. Since  
similar calculations to the ones in Sec.~\ref{a2g2}  show that for $2\pi^2k_bT/\hbar \omega_c 
 \:\raisebox{-0.4ex}{$\stackrel {>}{\sim}$}\: 1$  we can neglect the contribution to the zero harmonic from 
 the $I_{l,l}$-terms where $l \neq 0$, we conclude that $a_1$ for  is independent of the temperature for temperatures 
that are not too low. We have checked this independence against the exact result given in Eq.(\ref{alpha}) 
and found very good agreement. To obtain the dependence on $n_f$ and $\sigma$ we make the 
simplification 
\begin{equation} \tanh[K\sigma(v+u)]+\tanh[K\sigma(v-u)]v^{-1}\simeq\left\{
\begin{array}{ll} 0 & \mbox{if $|v|<|u|$}\\ \frac{2}{|v|} & \mbox{if $|v|>|u|$} 
\end{array} \right. \end{equation} 
which is a very good approximation since $K\gg1$. It is exact for $T=0$. The integral can be solved and 
we obtain:
\begin{equation} I_{0,0} \simeq \frac{4}{\sqrt{1+n_f/\sigma^2}}\ln \left( \sqrt{\frac{\sigma^2}{4n_f}+1}+
\sqrt{\frac{\sigma^2}{4n_f}+2}\right)
\simeq 2\ln(\frac{\sigma^2}{n_f}) \end{equation}
where we have assumed $\sigma^2 \gg 4n_f$. This  yields the result
 \begin{equation} \label{a1} a_1 \simeq \frac{V^2 \left(\frac{a_x}{l}\right) }
{4\pi L_x L_yl^2 \hbar \omega_ c}  \end{equation}
The expression for $a_1$ is independent of any spin effects for $\min(\sqrt{n_f},\sigma)\gg g$.
We have again checked the independence of $a_1$ on $n_f$, $\sigma$, and spin  against the exact result 
and we find very good agreement.

The dependence of $g_1$ on $n_f$, $\sigma$, and $T$ is determined by the integrals in Eq.(\ref{gint})
 for which $l_1-l_2+l_3-l_4=0$.  Again it turns out, that for 
$2\pi^2k_bT/\hbar \omega_c \: \raisebox{-0.4ex}{$\stackrel {>}{\sim}$}\:1 $ we can neglect the 
contribution to $g_1$ from the terms with $l_1-l_2+l_3-l_4=0$ and $\max(|l_1|,|l_2|,|l_3|,|l_4|)>0$.
 Using Eq.(\ref{matsusum}) we can rewrite the integral with $l_1=l_2=l_3=l_4=0$ as
\begin{eqnarray}
\int dx_1\!\ldots dx_4 \frac{\hbar\omega_c}{2} \left[\frac{1}{x_3-x_1}\left(
\frac{\tanh(Kx_1/2)}{(x_1+x_2)(x_1+x_4)}-\frac{\tanh(Kx_3/2)}{(x_3+x_2)(x_3+x_4)} \right)+
\right. & \nonumber \\ \left. \frac{1}{x_4-x_2} \left(\frac{\tanh(Kx_2/2)}{(x_1+x_2)(x_2+x_3)}-
\frac{\tanh(Kx_4/2)}{(x_4+x_1)
(x_4+x_3)} \right) \right]&\times \nonumber \\ e^{[(x_1-x_4)^2+(x_3-x_2)^2+(x_1-x_2)^2+
(x_3-x_4)^2]/8n_f}  e^{-2(x_1^2+x_2^2+x_3^2+x_4^2)/\sigma^2}
\Xi^{x_1+x_4,x_2+x_3}_{x_1+x_2,x_3+x_4} &
\end{eqnarray}
where $K=\hbar\omega_c/k_bT$ determines the  temperature dependence. Again for 
 $\min(\sqrt{n_f}, \sigma) \gg g$, $g_1$ will be independent of spin effects.  As in the case of $a_1$, it is 
fairly straightforward to see that since \mbox{$1/K \ll \min(\sqrt{n_f},\sigma)$}, the integral and therefore 
$g_1$  are independent of the temperature to a very good approximation. We have checked this 
independence against the exact result given in Eq.(\ref{gamma}) and find very good agreement. 
 
To make any progress in determining the dependence of $g_1$ on $\sigma$ and $n_f$ we need some 
 simple expression for $\Xi^{n_1+n_4,n_2+n_3}_{n_1+n_2,n_3+n_4}$. As a rough approximation we 
make the following simplification to:
\begin{equation} \Xi^{n_1+n_4,n_2+n_3}_{n_1+n_2,n_3+n_4}\sim (\delta_{n_1,n_3}+
\delta_{n_2,n_4})\frac{L_x L_y}{4\pi a_x^2} \end{equation}
This is based on the fact that $\chi_{j_1} ({\mathbf{k}} ) \chi^*_{j_2}( \mathbf{k} )$ in general is 
a complex number for  $j_1 \neq j_2$. When the $\mathbf{k}$-sum is performed the phase factor will
 change 'randomly' and make the sum approximately zero. Physically it corresponds to ignoring cases
 where electrons in four different Landau levels interact.
Using this simplification, Eq.~(\ref{Bapprox}) and the Poisson formula we get from  Eq.~\ref{gamma}
 the following integral determining the  dependence of $g_1$ on $n_f$ and $\sigma$:
\begin{equation}
 \label{poisson} \sum_{\omega'}\frac{k_bT}{n_f} \int dx dx_2 
dx_4 \frac{e^{-[(x-x_2)^2+(x-x_4)^4]/4n_f}}{(i\omega'-x)^2(i\omega'+x_2)(i\omega'+x_4)} 
 e^{-2(2x^2+x_2^2+x_4^2)/\sigma^2}=\sum_{\omega'}I_{\omega'}
\end{equation}
where $\omega'=\omega_{\nu}/\omega_c$. We have again assumed  
$2\pi^2k_bT/\hbar \omega_c \: \raisebox{-0.4ex}{$\stackrel {>}{\sim}$}\:1 $.  
Assume now that $8n_f \ll \sigma^2$. We approximate the integrals by:
\begin{equation}I_{\omega'} \simeq \frac{k_bT}{n_f}\int dx \frac{e^{-x^2/2n_f}}{(i\omega'-x)^2}
\int dx_2 \frac{e^{-x_2^2/4n_f}}{i\omega'+x_2}
 \int dx_4 \frac{e^{-x_4^2/4n_f}}{i\omega'+x_4} \end{equation}
We will now show, that in this approximation the sum of the integrals is  largely independent of $n_f$ and 
therefore $g_1 \propto 1/n_f$. The integral can  be solved and we get:
\begin{equation} I_{\omega'}=\frac{k_bT}{n_f^{3/2}}\left(-\frac{\pi|\omega'|}{\sqrt{n_f}}
e^{\omega'^2/2n_f}[1-\Phi(\frac{|\omega'|}{\sqrt{2n_f}})]+\sqrt{2\pi}\right)\pi^2
[1-\Phi(\frac{|\omega'|}{2\sqrt{n_f}})]^2e^{\omega'^2/2n_f} \end{equation}
where we again have $\Phi(x)\equiv \frac{2}{\sqrt{\pi}}\int_0^x dt e^{-t^2}$. Eq.(\ref{poisson}) can then be
 written on the form
\begin{equation}\frac{\hbar \omega_c}{2\pi n_f}\Delta x \sum_{x_n}\left(-\pi x_ne^{x_n^2/2}
[1-\Phi(\frac{x_n}{\sqrt{2}})]+\sqrt{2\pi}\right) \pi^2[1-\Phi(\frac{x_n}{2})]^2e^{x_n^2/2} 
\end{equation}
where $x_n=\frac{\Omega_N}{\omega_c\sqrt{n_f}}$ and $\Delta x=\frac{2\pi k_bT}
 {\hbar \omega_c \sqrt{n_f}}$. Since  $\Delta x \ll 1 $ we can approximate this sum by an integral and we
 therefore conclude that $g_1$ is independent of the temperature in agreement with the result above. 
Furthermore we obtain $g_1 \propto 1/n_f$ for $n_f$ large and $g_1$ independent of $\sigma$. When
 $\sigma^2 \gg 8n_f$ does not hold the calculation is the same as above. We just have to substitute 
$ 1/4n_f$ with $1/4n_f+2/\sigma^2$ in the integrals. The $1/n_f$ dependence coming from the 
$B_0^{j_1\ j_2}$ factors in Eq.~\ref{gamma} is unaltered and we still get that 
 $g_1 \propto 1/n_f$ for $\min(\sqrt{n_f},\sigma)$ large and that $g_1$ is independent of 
 $\sigma$ and the temperature. By calibrating $g_1$ through an exact evaluation based on 
Eq.(\ref{gamma}), we obtain:
\begin{equation} \label{g_1} g_1 \simeq \frac{V^4}{(L_xL_y)^3 l^2(\hbar\omega_c)^35.4n_f }
\end{equation} 
where $g_1$ is defined in 
Section~\ref{simple}. It should be noted that the dependence of $g_1$ on $n_f$ and $\sigma$ in the above 
expression is only approximate and rests on the various simplifications made.  We have tested the above 
expression against the exact result  and we find that the dependence on $n_f$ and $\sigma$ fits to an
 accuracy of 20\%.

\newpage
\begin{center}{Figure Captions}\end{center}
\bigskip
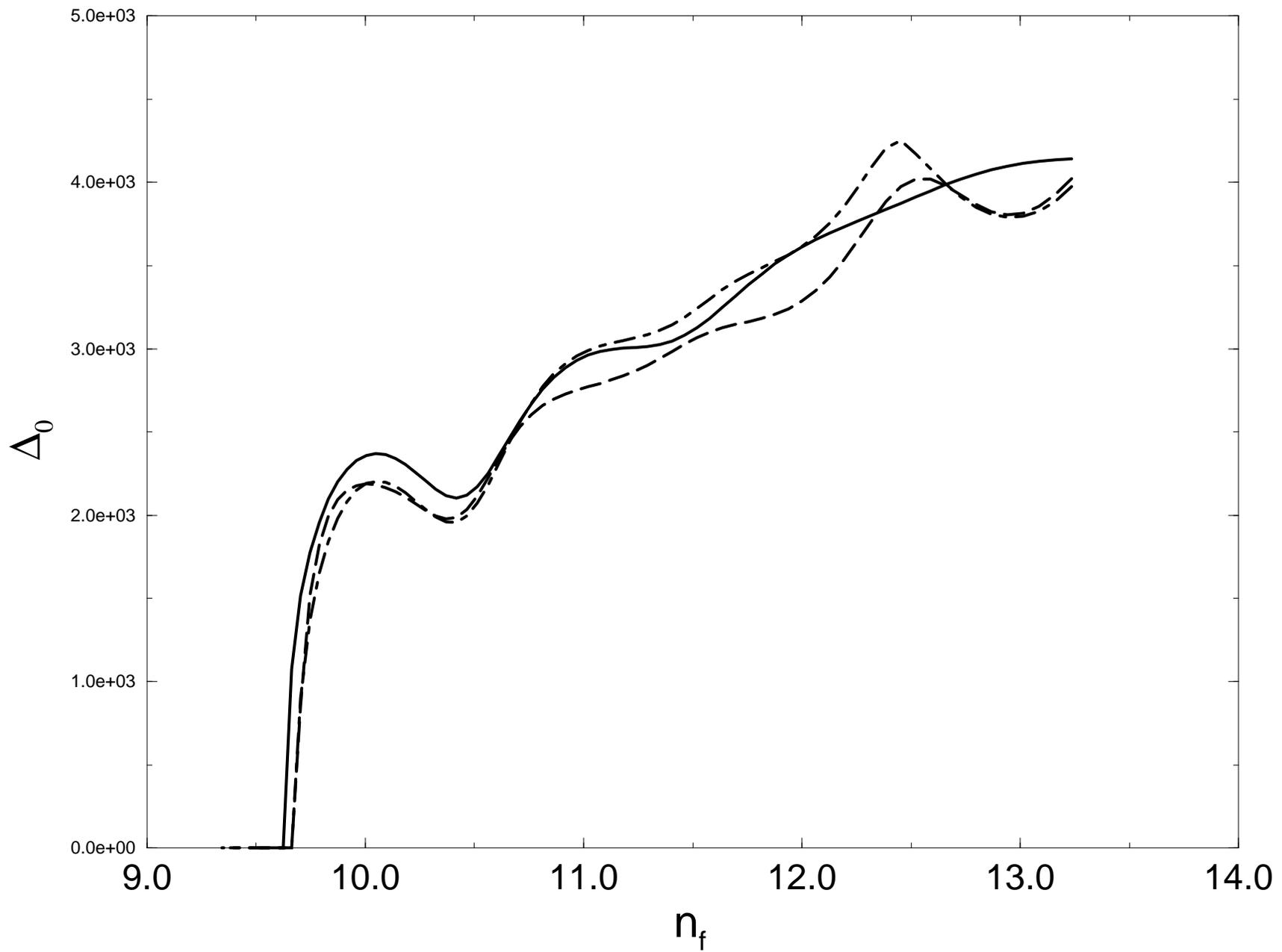
\noindent Fig.\ 1: The order parameter $\Delta_0$ vs $n_f$ calculated numerically (solid line), to fourth 
order in $\Delta_0$ (dashed line), and to eight order in $\Delta_0$ (dash-dot line).

\

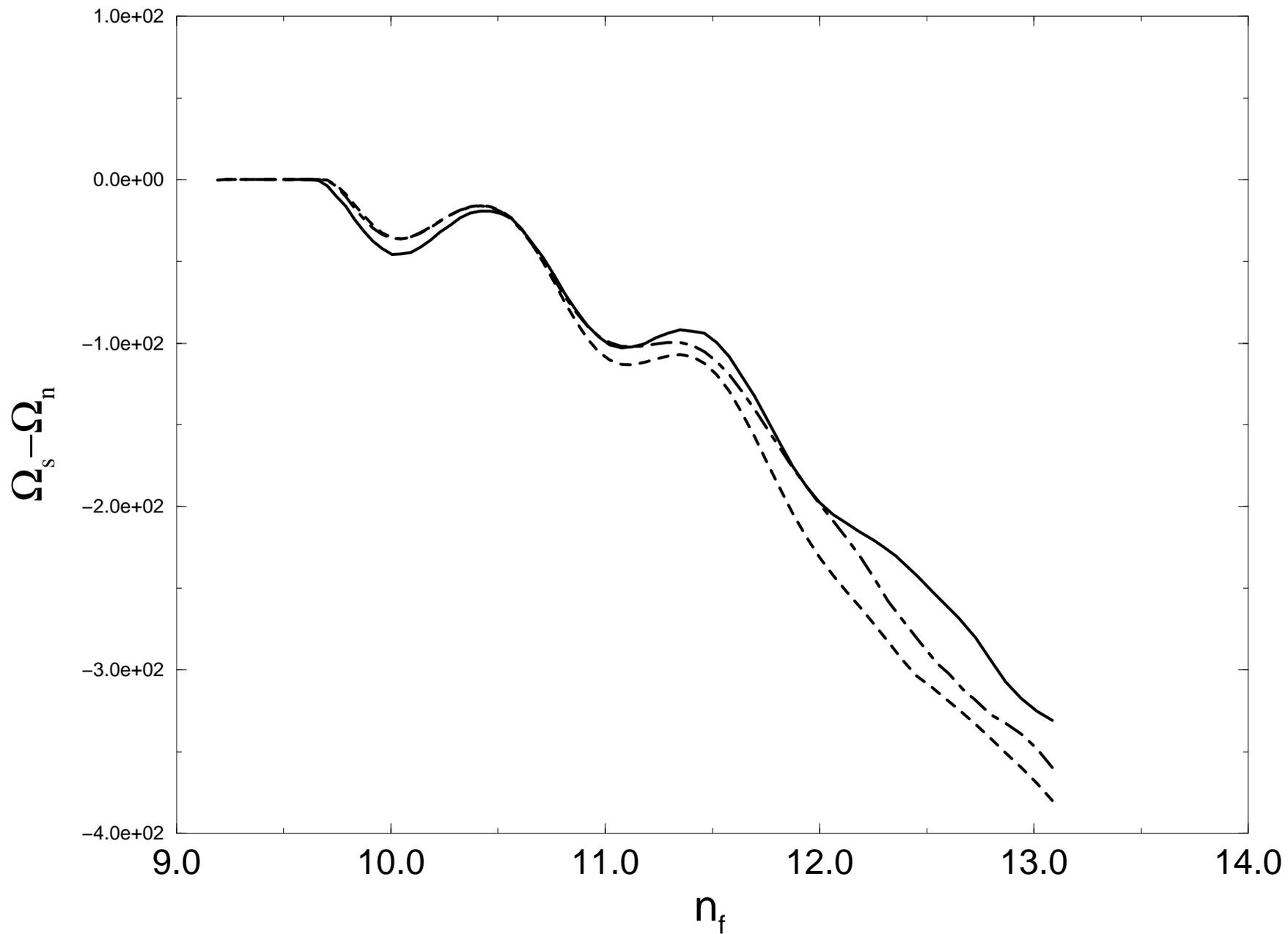
\noindent Fig.\ 2: The difference $\Omega_S-\Omega_N$ in the grand potential between the mixed state 
and the normal state. The solid line is a numerical calculation, the dashed line is fourth order perturbation 
theory, and the dash-dot line is eighth order perturbation theory.

\
 
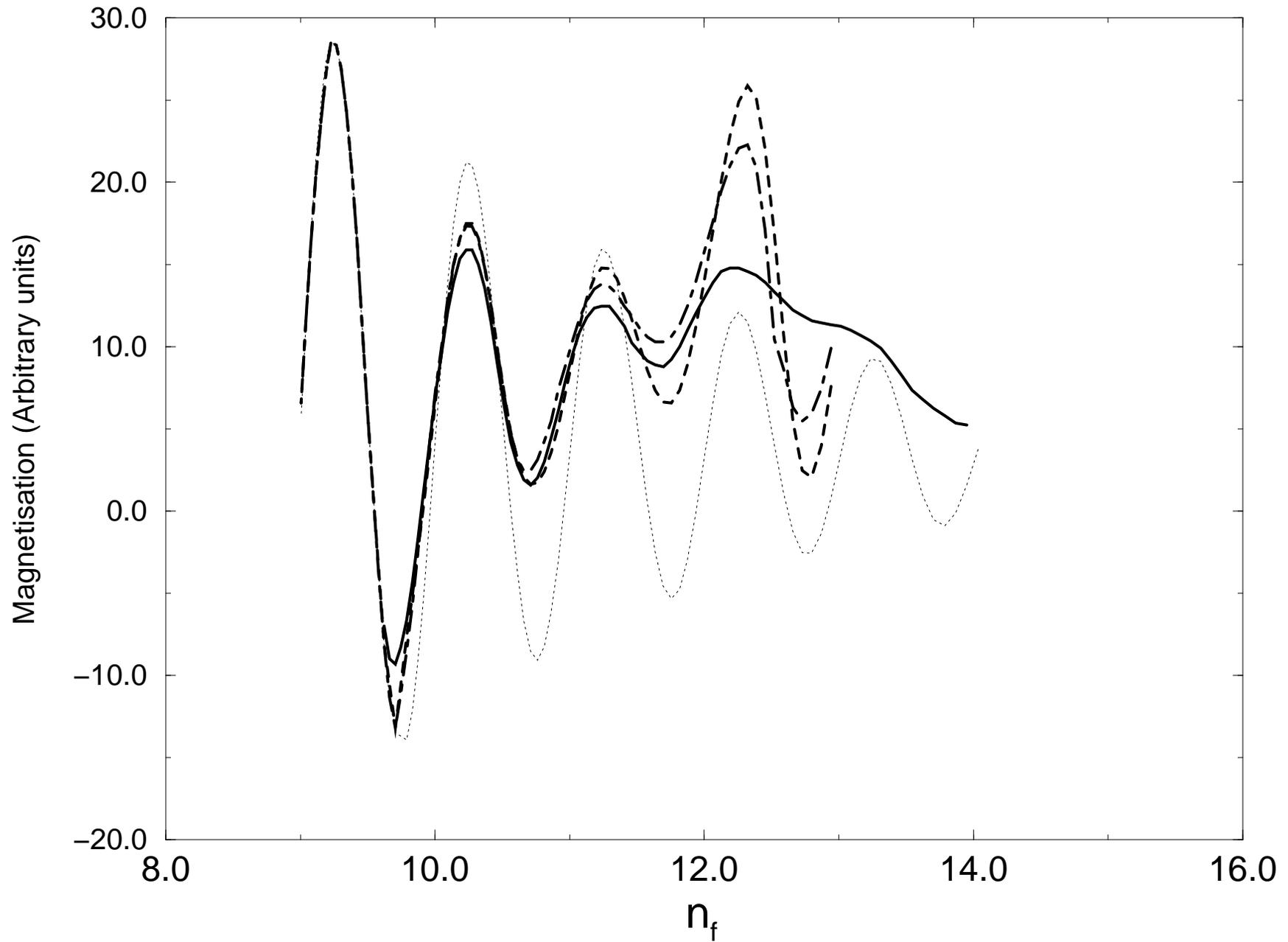
\noindent Fig.\ 3: The magnetization vs $n_f$. The solid line is a numerical calculation, the dashed line 
fourth order, the dash-dot line eight order, and the dotted line is the magnetization in the underlying 
normal state.

\

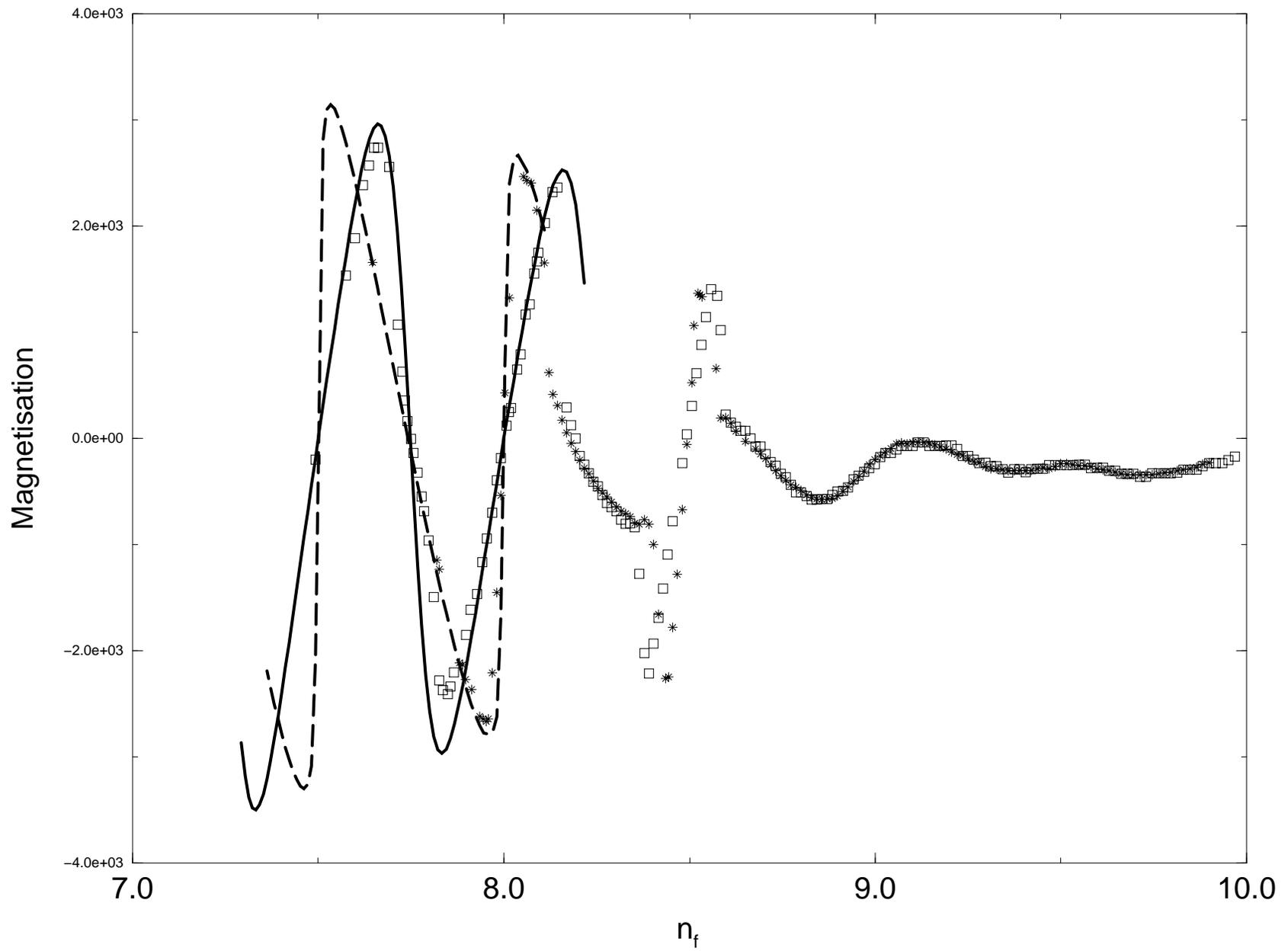
\noindent Fig.\ 4: The magnetization when the chemical potential is constant ($\Box$) and when the number
 of particles is constant ($\ast$) for a very low temperature. The solid and dashed lines are the normal 
state magnetization for fixed chemical potential and fixed number of particles respectively.

\

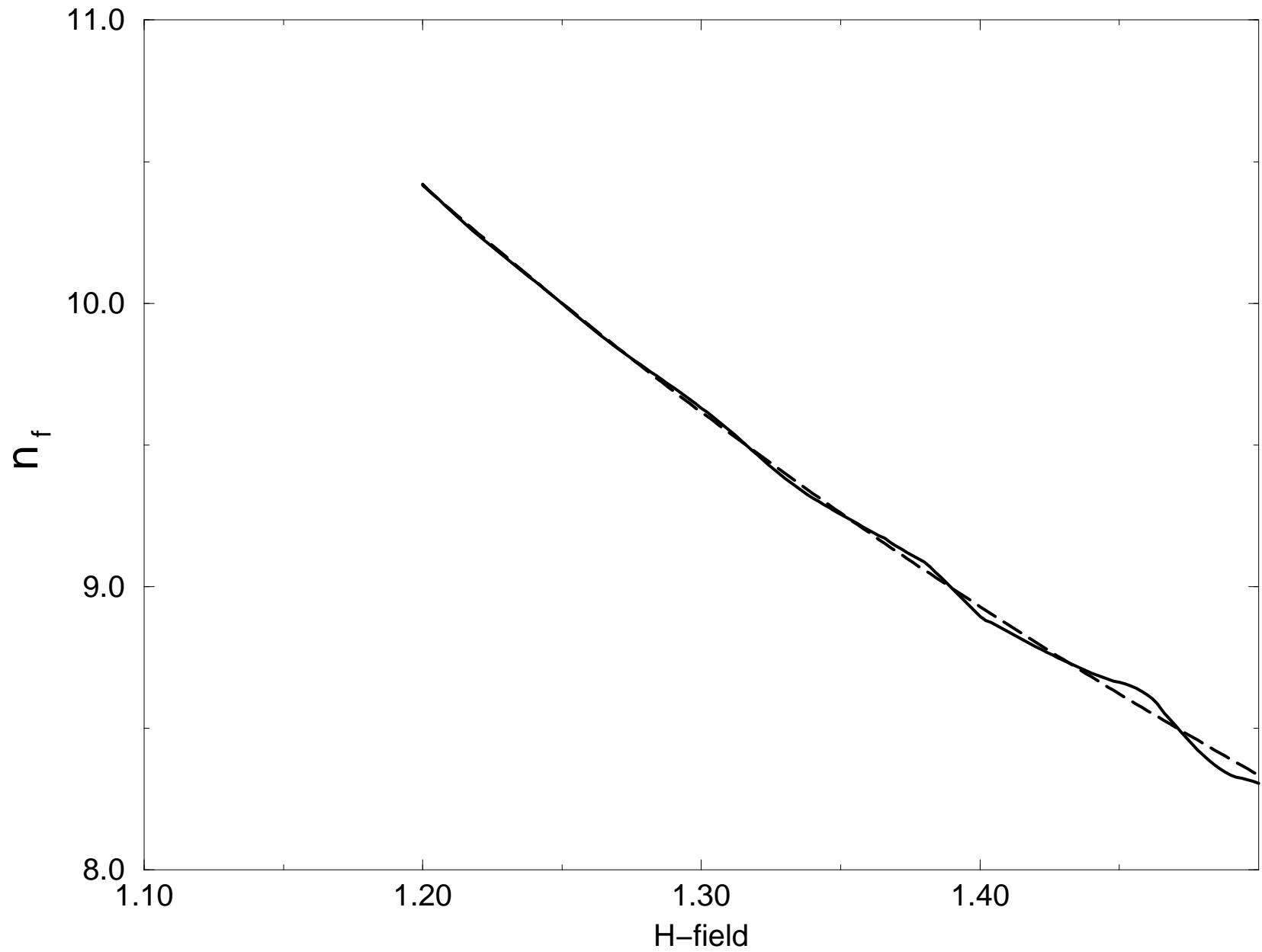
\noindent Fig.\ 5: $n_f$ as a function of the magnetic field for fixed chemical potential (dashed line)
 and fixed number of particles (solid line).

\

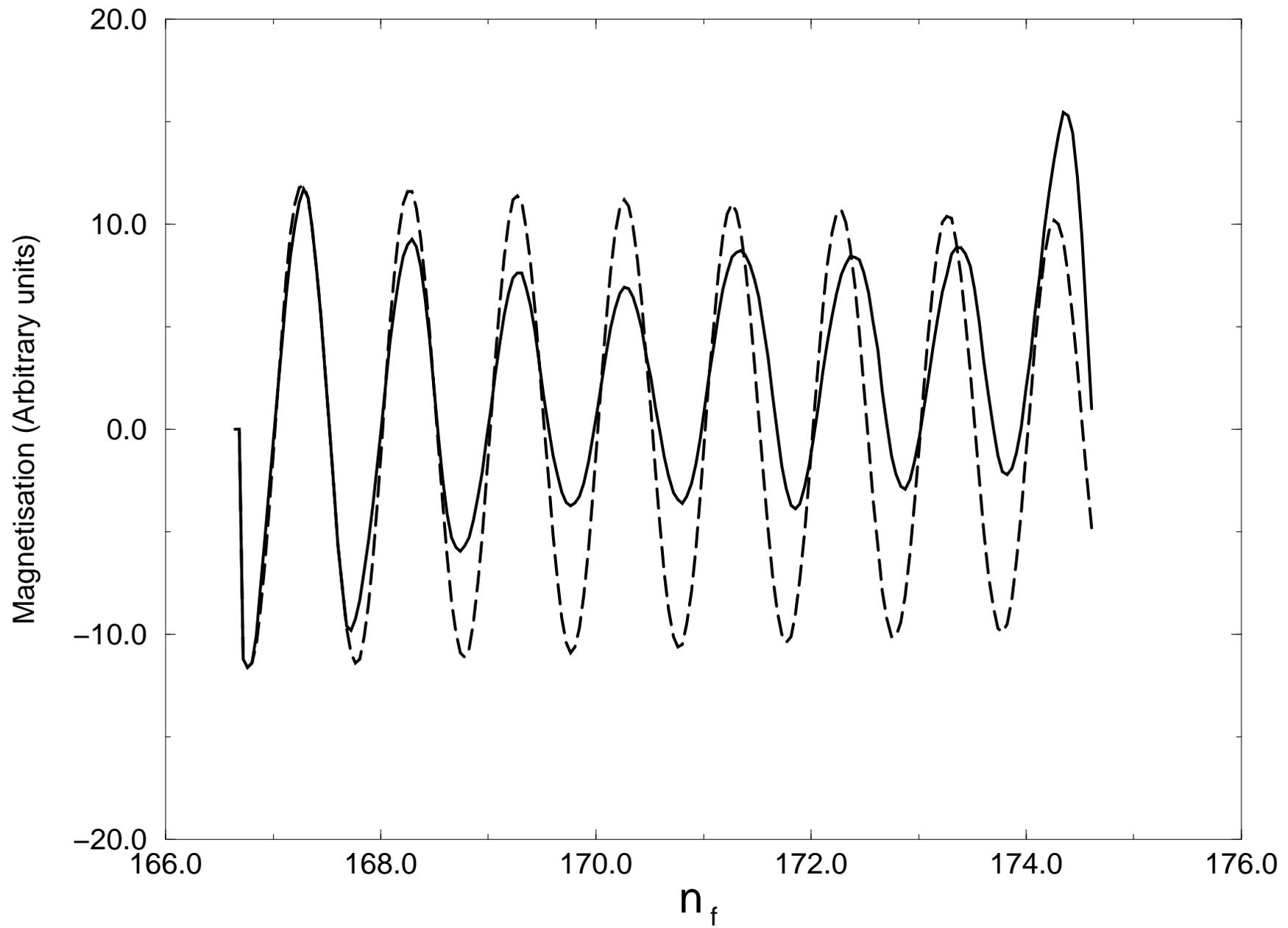
\noindent Fig.\ 6: The magnetization vs $n_f$ in the mixed state (solid line) and in the underlying normal
state (dashed line).

\

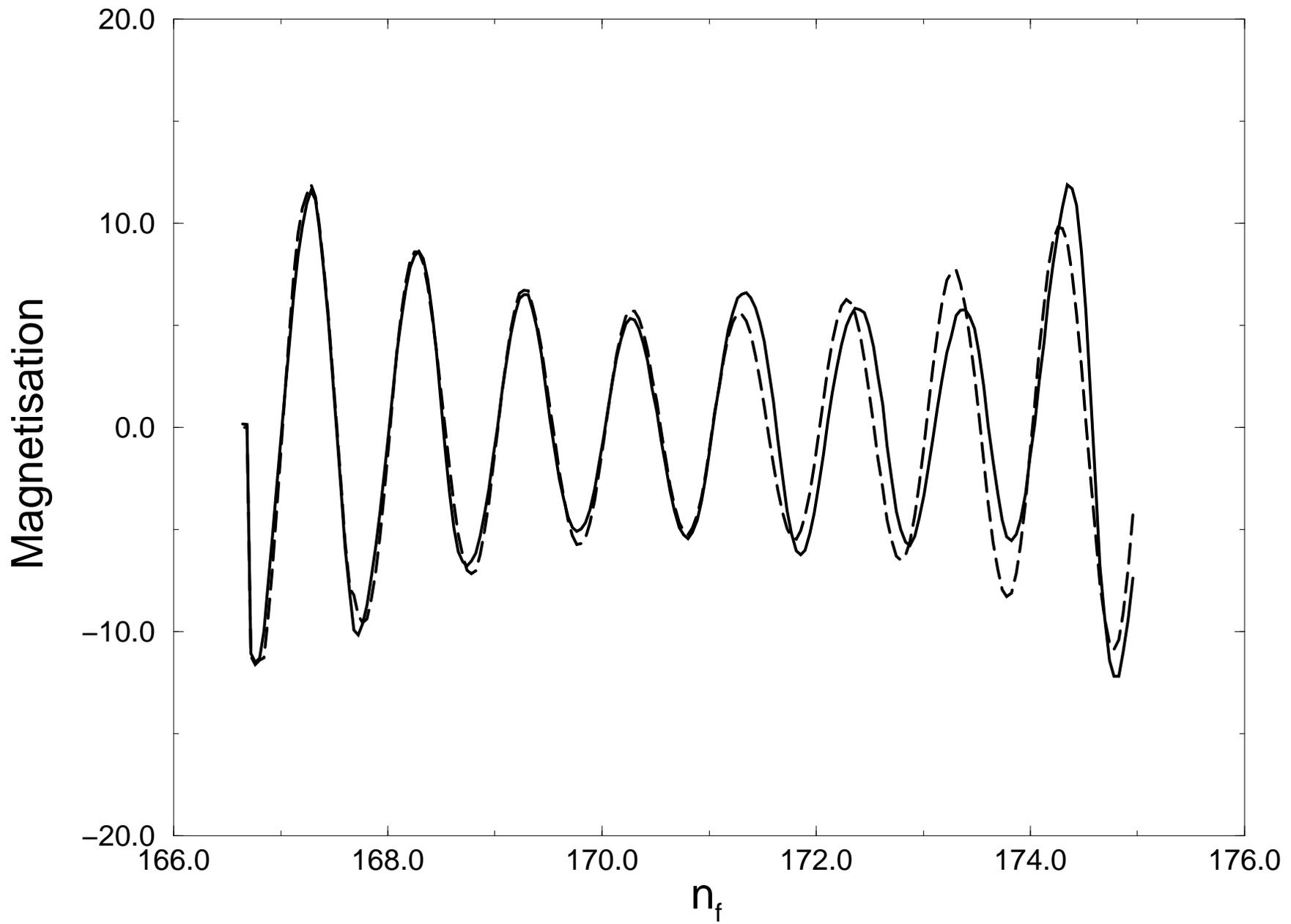
\noindent Fig.\ 7: The magnetization vs $n_f$ in the mixed state. The solid line is the first harmonic of the 
perturbative calculation and the dashed line is obtained from Eq.\ (\ref{1harms}). 

\

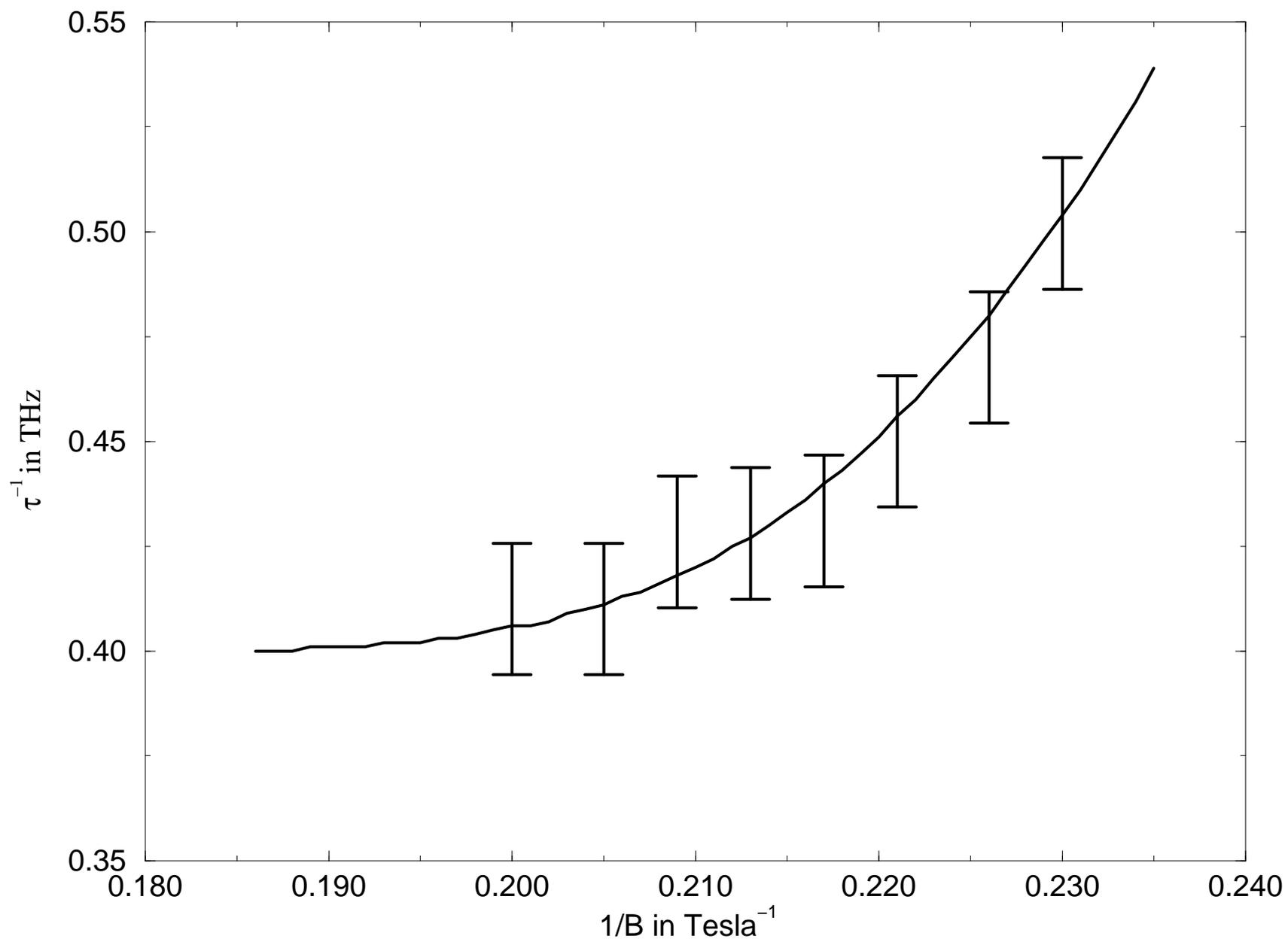
\noindent Fig.\ 8: $\tau^{-1}$ as a function of $1/B$. The solid line is the theoretical prediction and the 
bars are the experimental data.~\cite{Caulfield}
\end{document}